\documentclass{iopjournal}
\usepackage{graphicx,epsfig}
\usepackage{amsmath,amssymb,wasysym}
\usepackage{dsfont} 
\usepackage{color}

\newcommand{\op}[1]{\hat{#1}}
\newcommand{\normV}[1]{||{#1}||_2}

\newcommand{\normspec}[1]{||{#1}||_{\text{2}}}

\newcommand{\ident}{\mathds{1}}
\newcommand{\C}{\mathbb{C}}

\newcommand{\R}{\mathbb{R}}

\newcommand{\rem}[1]{}

\newcommand{\imagc}[1]{\text{Im}\,#1}

\newcommand{\image}[1]{\text{Im}\left(#1\right)}

\newcommand{\realc}[1]{\text{Re}\,#1}
\newcommand{\oHami}{\op{\Gamma}}

\newcommand{\bra}[1]{\langle#1|}
\newcommand{\ket}[1]{|#1\rangle}
\newcommand{\transpose}{{\text{T}}}
\newcommand{\braket}[2]{\langle#1|#2\rangle}

\newcommand{\rca}{\xi}
\newcommand{\Hperturbed}{\op{H}}
\newcommand{\Hs}{\op{H}_0}
\newcommand{\Hp}{\op{H}_1}

\newcommand{\order}{n}

\newcommand{\ev}{\omega}

\newcommand{\evEP}{\ev_{\text{EP}}}
\newcommand{\evDP}{\ev_{\text{DP}}}

\newcommand{\ind}{l}
\newcommand{\PF}{K}
\newcommand{\projector}{\op{P}}
\newcommand{\nilpotent}{\op{N}}

\newcommand{\Smatrix}{\op{S}}
\newcommand{\GFs}{\op{G}_0}
\newcommand{\GF}{\op{G}}
\newcommand{\coupling}{\op{W}}
\newcommand{\FI}{{\cal{I}}}
\newcommand{\FIred}{{\cal{I}}^\text{red}}
\newcommand{\FImax}{{\cal{I}}^\text{max}}
\newcommand{\FIavg}{{\cal{I}}^\text{avg}}
\newcommand{\FImod}{{\cal{I}}^\text{mod}}
\newcommand{\parameter}{\theta}
\newcommand{\incoming}{u^{\text{in}}}
\newcommand{\outgoing}{u^{\text{out}}}
\newcommand{\intensity}{\bar{n}^{\text{in}}}
\newcommand{\Hsystem}{\op{H}_{\text{sys}}}
\newcommand{\trails}{m}
\newcommand{\WS}{\op{Q}}
\newcommand{\DOS}{\rho}
\newcommand{\Qev}{u}

\newcommand{\HL}[1]{#1}

\begin{document}
\articletype{Paper} 

\title{Fundamental Limits of Non-Hermitian Sensing from Quantum Fisher Information}
\author{Jan Wiersig$^*$}

\email{jan.wiersig@ovgu.de}

\affil{$^1$Otto von Guericke University Magdeburg, Faculty of Natural Sciences, Institute of Physics, Universit{\"a}tsplatz 2, 39106 Magdeburg, Germany} 

\author{Stefan Rotter$^*$}

\let\thefootnote\relax\footnotetext{*Corresponding author}

\email{stefan.rotter@tuwien.ac.at}

\affil{$^2$Vienna University of Technology (TU Wien), Institute for Theoretical Physics, Wiedner Hauptstra\ss e 8-10/136, A-1040 Vienna, Austria} 

\begin{abstract}
Exceptional points (EPs) exhibit strongly enhanced spectral responses and are therefore promising candidates for sensing applications. Whether these non-Hermitian degeneracies provide a genuine advantage in the quantum regime has been the subject of ongoing debate. Here, we address this issue within a scattering-matrix formalism for sensing with coherent light, which allows the quantum Fisher information (QFI) to be evaluated directly from experimentally accessible scattering data without introducing additional noise channels beyond those inherent to the scattering process. We analyze both nondegenerate and degenerate scattering-matrix poles, including EPs of arbitrary order, and show that the QFI per incoming photon flux is bounded and controlled by three key factors: the decay rate of the resonant mode, the strength of the spectral response associated with non-normality, and the adjustment between the scattering states and the information source. For spatially localized perturbations, this implies that the maximal QFI achievable by wavefront shaping is fully determined by the local density of states at the perturbation site. Within this framework, we demonstrate that EPs can enhance the QFI compared to isolated modes or diabolic points with identical decay rates, and that the QFI can be further increased by moving away from the EP toward parameter regimes where non-Hermitian linewidth splitting reduces the decay rate of one mode. We further show that sufficiently small additional internal losses do not alter this overall picture, thereby providing a unified and experimentally relevant perspective on the design of quantum-limited non-Hermitian sensors.
\end{abstract}

\section{Introduction}
The physics on non-Hermitian systems, i.e., systems with gain and loss, has attracted considerable interest in the recent years. The non-Hermitian effects are most pronounced near and at non-Hermitian degeneracies called exceptional points (EPs). At an EP of order~$\order$ exactly~$\order$ eigenenergies and the corresponding energy eigenstates  coalesce~\cite{Kato66,Heiss00,Berry04,Heiss04,MA19}. This is in contrast to a conventional degeneracy (diabolic point, DP) where only eigenenergies coalesce while the eigenstates can be chosen to be orthogonal. EPs require both non-Hermiticity of the Hamiltonian, $\op{H} \neq \op{H}^\dagger$, and nonnormality, i.e., $[\op{H},\op{H}^\dagger] \neq 0$, and they have enabled a wide range of phenomena in 
optics and photonics~\cite{MA19}, such as loss-induced suppression of lasing~\cite{POR14}, mode discrimination in multimode lasing~\cite{HMH14}, orbital angular momentum microlasers~\cite{MZS16}, unidirectional lasing~\cite{POL16}, mode conversion~\cite{XMJ16,DMB16}, circularly-polarized light sources~\cite{RMS17}, chiral perfect absorption~\cite{SHR19}, optical amplifiers with improved gain-bandwidth product~\cite{ZOE20}, subwavelength control of light transport~\cite{XLJ23}, and as a resource for hardware encryption~\cite{YZZ23}.  

A notably relevant application is that of sensors based on EPs~\cite{Wiersig14b,Wiersig16}. Perturbing a Hamiltonian $\Hs$ with an EP by a generic perturbation $\Hp$ with perturbation strength~$\varepsilon$
\begin{equation}\label{eq:Hp}
\Hperturbed = \Hs+\varepsilon\Hp
\end{equation}
gives rise to an energy splitting proportional to the $\order$th root of~$\varepsilon$~\cite{Kato66}. For sufficiently small~$|\varepsilon|$ this exceeds the linear scaling observed near a DP or at an isolated mode. The strength of the response to small generic perturbations can be characterized by a single quantity, the spectral response strength~$\rca$~\cite{Wiersig22,Wiersig23b}. A large value of $\rca$ signals a significant spectral response. 

There have been several experimental realizations of EP-based sensors~\cite{COZ17,HHW17,XLK19,LLS19,LLS19,HSC19,PNC20,KCE22}, some with practical applications, such as ring laser gyroscopes~\cite{HSC19}, implantable microsensors in a rat abdomen~\cite{DLY19}, all-optical bending sensors~\cite{LCL23}, detection of weathering in the Yungang Grottoes~\cite{DDS25}, multiplexed wireless sensing systems for wearable applications~\cite{YZY25}. Still, there is an ongoing debate regarding whether EP-based sensors truly offer advantages over conventional sensors in terms of signal-to-noise ratio, particularly within the quantum regime~\cite{Wiersig20b,Wiersig20c}. The first experiment to observe the quantum limitations of EP-based sensors was conducted by Wang and colleagues using Brillouin laser gyroscopes~\cite{WLY20}. A critical observation was that the increasing enhancement in frequency splitting of the gyroscope is exactly counterbalanced by a corresponding broadening of the laser linewidths. Conversely, other studies have demonstrated an enhanced signal-to-noise ratio in non-Hermitian sensors exploiting a transmission peak degeneracy (TPD) near an EP~\cite{KCE22,XML24}. 

The debate also encompasses various theoretical contributions~\cite{Langbein18,LC18,ZSH19,CJL19,DMA22}, which suggest that frequency splittings are not measured directly; rather, they are derived from field or intensity measurements taken at different frequencies. These  approaches consider the so-called weak dispersive limit, a condition in which frequency shifts are significantly smaller than the spectral linewidths.
Basic calculations have demonstrated that the changes in the intensity have the same scaling
for both DPs and EPs~\cite{Langbein18}. Thus, parametrically, the response at an EP is not greater than at a DP, although the absolute magnitude of the response at an EP can still be larger~\cite{Sunada18,Wiersig22}.

Most of the theoretical approaches build on the quantum Fisher information (QFI), a general concept underlying high precision measurements of physical parameters. The QFI establishes a limit of precision of any unbiased estimator according to the quantum Cram\'er-Rao lower bound for the standard deviation~$\delta\parameter$~\cite{BC94}
\begin{equation}\label{eq:CramerRao}
\delta\parameter \geq \frac{1}{\sqrt{\trails\FI_\parameter}}
\end{equation}
where $\FI_\parameter$ is the QFI for the estimation of a single parameter $\parameter$ and $\trails$ represents the number of independent experimental trials used to derive the estimate. The Fisher information is directly related to the signal-to-noise ratio~\cite{LC18} and, to our knowledge, was first considered in the context of non-Hermitian systems in Ref.~\cite{BG13}.

Within the QFI framework, several papers have explored sensors operating at second-order EPs, primarily considering Gaussian quantum noise.
Reference~\cite{LC18} asserts that for reciprocal systems, sensors based on EPs do not offer an advantage regarding the quantum-limited signal-to-noise ratio. A recent experiment on superconducting qubits claims to confirm this assessment~\cite{AAK25}.
In contrast, other studies revealed that EP-based sensing using an amplifier near its lasing threshold can be advantageous also in the quantum regime, as the QFI exhibits a faster divergence speed when approaching the threshold~\cite{ZSH19,ASF23,NCK23}.
However, this analysis, relies on a linearization that could be problematic near the laser threshold~\cite{CJL19}.
For a microring coupled to a waveguide with partial mirror an enhanced QFI for coherent states and NOON states has been predicted recently~\cite{CKW26}. 
Finally, one argument is that, since any non-Hermitian system can be embedded within a Hermitian one, non-Hermitian sensors cannot surpass Hermitian sensors in performance~\cite{DWC23}. However, this reasoning overlooks a practical issue, namely that the corresponding Hermitian embedding may be too large to be fully controllable in experiments constrained by finite resources~\cite{ZLX25}.
\HL{For the special class of pseudo-Hermitian systems, a covariant formulation of the QFI was introduced, enabling the identification of optimal projections that saturate the classical Fisher information~\cite{ANO26}.
All the systems mentioned above function in the linear regime; those operating in the non-linear regime come with their own unique challenges~\cite{ZC25}.}

Recently, Bouchet \emph{et al.} have shown that for scattering measurements with coherent light and Poissonian noise statistics (which includes Gaussian noise as limiting case), the QFI can be fully extracted from the output signal using a homodyne detection scheme and it can be straightforwardly evaluated with the classical scattering matrix~\cite{BRM21}. At a given input frequency $\ev$ this matrix $\Smatrix(\ev)$ maps incoming states $\ket{\incoming}$ to outgoing states $\ket{\outgoing} = \Smatrix(\ev)\ket{\incoming}$ ($\ket{\incoming}$ and $\ket{\outgoing}$ are assumed to be energy flux-normalized). Based on the experimentally available $\Smatrix(\ev)$~\cite{PLC10,RG17,CMR22}, the incoming-state dependent QFI can be written as
\begin{equation}\label{eq:FIdef}
\FI_\parameter(\incoming) = 4\bra{\incoming}(\partial_\parameter\Smatrix)^\dagger\partial_\parameter\Smatrix\ket{\incoming}/(\hbar \omega) .
\end{equation}
Note that $\partial_\parameter\Smatrix\ket{\incoming}$ quantifies how the outgoing state changes with the parameter $\parameter$ and the factor $1/(\hbar \omega)$ translates the energy flux to photon flux as relevant for the Poissonian noise statistics in the coherent light beams assumed here. 
In these units, the QFI defined in Eq.~(\ref{eq:FIdef}) can be considered as a QFI per unit time~\cite{HRR24}.

The aim of this paper is to use the approach of Bouchet \emph{et al.} as an easy-to-use scattering formalism for studying 
\HL{intrinsic QFI bounds and scaling relations in non-Hermitian sensing. To do so, we derive upper bounds for the QFI, which correspond to lower bounds on the achievable estimation uncertainty through the quantum Cramér–Rao relation~(\ref{eq:CramerRao})} in the presence of both degenerate and nondegenerate modes. We demonstrate that in a system at an EP, its spectral response strength determines the QFI, whereas for a DP and for an isolated mode the corresponding Petermann factor is of relevance.
\HL{Beyond addressing whether EPs can enhance quantum-limited sensing, the central aim of this work is to identify the minimal physical ingredients that control the extraction of Fisher information in non-Hermitian scattering systems, and to clarify how these ingredients reconcile the seemingly contradictory conclusions reported in the recent literature. Our analysis focuses on the intrinsic QFI associated with coherent scattering measurements under shot-noise-limited conditions, without incorporating additional technical noise sources beyond those inherent to the scattering process itself. In this sense, our aim is not to provide a complete description of experimentally achievable sensor performance under arbitrary noise conditions, but rather to identify the intrinsic mechanisms governing the QFI in non-Hermitian scattering systems.}

The outline of the paper is as follows. In Sect.~\ref{sec:preliminaries} the basics on Green's functions for non-Hermitian systems, the spectral response strength of EPs, and  the scattering approach to the QFI are introduced. Our theory is developed in Sect.~\ref{sec:FIDPEP}. Section~\ref{sec:examples} provides some waveguide-microring systems as illustrative examples. The important issue of internal losses is discussed in Sect.~\ref{sec:losses}. Finally, a summary and outlook is given in Sect.~\ref{sec:conclusion}.

\section{Preliminaries}
\label{sec:preliminaries}
We consider a generic scattering-based sensing setup in which an open system is probed by coherent light and a system parameter $\parameter$ is to be estimated from the outgoing fields. The parameter dependence enters through a perturbation of the internal Hamiltonian, for example due to a local frequency shift or induced scattering. The system is described within a scattering-matrix formalism, which naturally accounts for openness and non-Hermiticity arising from coupling to external channels. Our aim is to characterize the QFI associated with such measurements and to identify the factors governing its magnitude in the presence of isolated modes, diabolic points, and exceptional points. To this end, we briefly review the required properties of non-Hermitian Green’s functions and spectral response strengths, and then connect them to the scattering-matrix expression for the QFI.

\subsection{Green's functions of non-Hermitian systems}
\label{sec:Gxi}
In the following, we set $\hbar = 1$, so the numerical value of energy~$E$ is equal to the numerical value of frequency~$\omega$.
We start with the general expansion of a non-Hermitian $m\times m$ Hamiltonian
\begin{equation}\label{eq:Hexpand}
	\op{H} = \sum_\ind (\ev_\ind\projector_\ind+ \nilpotent_\ind)
\end{equation}
with~$\ind = 1,2,\ldots$ running over the considered part of the point spectrum $\ev_\ind$ including isolated eigenfrequencies and EPs. The operators~$\projector_\ind$ are projectors onto the generalized eigenspaces of the corresponding eigenvalues~$\ev_\ind$ with
\begin{equation}\label{eq:pp}
	\projector_j\projector_\ind = \delta_{j\ind}\projector_\ind 
\end{equation}
and Kronecker delta $\delta_{j\ind}$. The $\projector_\ind$ are not orthogonal projectors in general, i.e., $\projector_\ind \neq \projector^\dagger_\ind$.   
The operators~$\nilpotent_\ind$ for a given EP of order $\order_\ind \geq 2$ are nilpotent operators of index~$\order_\ind$; hence, $\nilpotent_\ind^{\order_\ind} = 0$ but $\nilpotent_\ind^{\order_\ind-1} \neq 0$. It holds that
\begin{equation}\label{eq:pn}
	\projector_\ind \nilpotent_\ind = \nilpotent_\ind\projector_\ind = \nilpotent_\ind  .
\end{equation}
The $m\times m$ Green's function of the Hamiltonian~$\Hs$ can be expanded as~\cite{Kato66}
\begin{equation}\label{eq:GKato}
	\GF(\ev) = \sum_\ind\left[\frac{\projector_\ind}{\ev-\ev_\ind} + \sum_{k=2}^{\order_\ind} \frac{\nilpotent_\ind^{k-1}}{(\ev-\ev_\ind)^k}\right] .
\end{equation}
\subsection{The spectral response strength of EPs}
\label{sec:xi}
For a frequency $\ev$ close to an EP frequency $\evEP = \ev_{\ind}$, the contribution of the Green's function with $k=\order_{\ind} = \order$ is the dominant one. 
Using this  Green's function with $\nilpotent = \nilpotent_\ind$
\begin{equation}\label{eq:GatEP}
\GFs(\ev) = \frac{\nilpotent^{\order-1}}{(\ev-\evEP)^{\order}} ,
\end{equation}
it has been shown in Ref.~\cite{Wiersig23b} that 
for a generic perturbation~(\ref{eq:Hp}) with small perturbation strength $\varepsilon$, the following inequality holds for the split eigenvalues $\tilde{\ev}_j$, $j = 1,\ldots,\order$, of the perturbed Hamiltonian
\begin{equation}\label{eq:ub}
	|\tilde{\ev}_j-\evEP|^{\order} \leq |\varepsilon| \normspec{\Hp}\,\rca 
\end{equation}
with the spectral norm of a matrix $\op{A}$~\cite{Johnston21}
\begin{equation}\label{eq:defspn}
\normspec{\op{A}} := \max_{\normV{\psi} = 1}\normV{\op{A}\psi} .
\end{equation}
We use the notation $\normspec{\cdot}$ both for the spectral norm of a matrix and the 2-norm $\normV{\psi} = \sqrt{\braket{\psi}{\psi}}$ of a vector~$\ket{\psi}$.
The spectral norm also enters the so-called spectral response strength~\cite{Wiersig22,Wiersig23b}
\begin{equation}\label{eq:rcagen}
	\rca = \normspec{\nilpotent^{\order-1}}  .
\end{equation}
The spectral response strength is a measure of non-normality associated with the EP. It quantifies how strongly a non-Hermitian system with an EP can react to a generic perturbation as specified in Eq.~(\ref{eq:ub}). This quantity has been analyzed and computed for numerous systems~\cite{Wiersig22b,Wiersig22c}, with its connection to the Petermann factor explored in previous studies~\cite{Wiersig23,KWS25}. A recent development includes a methodology for computing the spectral response strength directly through numerical wave simulations~\cite{KW25}.

\subsection{Fisher information in the scattering setup}
\label{sec:FIscattering}
In the following, we express the parameter-dependence of the scattering matrix ($\partial_\parameter\Smatrix$) as appearing in Eq.~(\ref{eq:FIdef}) for the QFI through the Green's function. Following the supplementary information of Ref.~\cite{HKB20}, our starting point is the Mahaux-Weidenmüller formula for the $M \times M$ scattering matrix~\cite{MW69}
\begin{equation}\label{eq:MW}
\Smatrix(\ev) = \ident_M-2i\coupling^\dagger \GF(\ev)\coupling 
\end{equation}
with the $N \times N$ Green's function 
\begin{equation}\label{eq:G}
	\GF(\ev) = (\ev\ident_N-\Hperturbed)^{-1}
\end{equation}
of the  $N \times N$ effective Hamiltonian
\begin{equation}\label{eq:Heff}
\Hperturbed = \Hsystem -i\coupling\coupling^\dagger 
\end{equation}
and the $M \times M$ ($N \times N$) identity matrix $\ident_M$ ($\ident_N$).
The $N \times M$-matrix $\coupling$ describes the coupling of the $M$ incoming scattering channels to the $N$ internal modes of the system, and  the $M \times N$-matrix $\coupling^\dagger$  describes the coupling of the $N$ internal modes to the $M$ outgoing scattering channels. $\Hsystem$ is the Hamiltonian of the uncoupled system, which is assumed to be Hermitian. In this case, the scattering matrix in Eq.~(\ref{eq:MW}) is unitary.

Rewriting Eq.~(\ref{eq:G}) 
\begin{equation}\label{eq:EHGF}
	(\ev\ident_N-\Hperturbed)\GF(\ev) = \ident_N 
\end{equation}
and taking the derivative with respect to the parameter~$\parameter$ on both sides yields
\begin{equation}
	\partial_\parameter\GF(\ev) = \GF(\ev)(\partial_\parameter \Hperturbed)\GF(\ev) .
\end{equation}
Using this equation and Eq.~(\ref{eq:MW}) gives
\begin{equation}\label{eq:dsm}
	\partial_\parameter \Smatrix = -2i\coupling^\dagger \GF(\ev) (\partial_\parameter \Hperturbed) \GF(\ev) \coupling ,
\end{equation}
assuming that $\coupling$ does not depend on $\parameter$. The latter assumption together with $\Hsystem$ being Hermitian and Eq.~(\ref{eq:Heff}) implies that $\partial_\parameter \Hperturbed$ is also Hermitian.

As in the QFI only $(\partial_\parameter\Smatrix)^\dagger\partial_\parameter\Smatrix$ enters, one can write Eq.~(\ref{eq:FIdef}) with unitary $\Smatrix$ also as 
\begin{equation}\label{eq:FIdefQ}
	\FI_\parameter(\incoming) = 4\bra{\incoming}Q_\parameter^2 \ket{\incoming}/(\hbar\omega).
\end{equation}
The Hermitian operator
\begin{equation}\label{eq:Qdef}
\WS_\parameter := -i\Smatrix^\dagger  \partial_\parameter\Smatrix 
\end{equation}
can be considered as a generalized Wigner-Smith operator~\cite{ABB17,HKB20}.
Multiplying Eq.~(\ref{eq:EHGF}) from the left by~$\GF^\dagger$ and
subtracting the adjoint of the resulting equation gives
\begin{equation}\label{eq:Gdown1}
\GF^\dagger - \GF = 2i\GF^\dagger\coupling\coupling^\dagger\GF ,
\end{equation}
taking advantage of the Hermiticity of $\Hsystem$.
Using this equation together with definition~(\ref{eq:Qdef}) and Eq.~(\ref{eq:dsm}) one obtains
\begin{equation}\label{eq:Qscatt}
\WS_\parameter  = -2\coupling^\dagger\GF^\dagger(\partial_\parameter \Hperturbed) \GF\coupling .
\end{equation}

\subsection{Fisher information for perturbed Hamiltonians}
\label{sec:FIperturbations}
We now apply the theory from the previous section to \HL{a Hamiltonian that we write in the form of a  perturbed system described by Eq.~(\ref{eq:Hp}). Here, $\Hs$ denotes the unperturbed Hamiltonian of interest (which may later exhibit an EP, DP, or isolated mode) and it is taken to be located at $\varepsilon = 0$. The Hermitian perturbation $\Hp$ and the real-valued perturbation parameter $\varepsilon$ describe the effect of the observable parameter~$\parameter$.} 
The unperturbed Hamiltonian $\Hs$ is non-Hermitian due to the coupling to the scattering channels according to Eq.~(\ref{eq:Heff}). With the Green's function of the unperturbed Hamiltonian $\Hs$
\begin{equation}\label{eq:G0}
\GFs(\ev) = (\ev\ident_N-\Hs)^{-1}
\end{equation}
and Eq.~(\ref{eq:Qscatt}), we can write
\begin{equation}\label{eq:QH1}
\WS_\varepsilon = -2\coupling^\dagger \GFs^\dagger(\ev) \Hp\GFs(\ev) \coupling .
\end{equation}
With the spectral norm in Eq.~(\ref{eq:defspn}) we can write for the QFI in Eq.~(\ref{eq:FIdefQ})
\begin{equation}\label{eq:FIuin}
\FI_\varepsilon(\incoming) = 4\normV{\WS_\varepsilon \incoming}^2/(\hbar\omega) \leq 4\normspec{\WS_\varepsilon}^2\,\intensity ,
\end{equation}
with the incoming photon flux $\intensity := \normV{\incoming}^2/(\hbar\omega)$. Inserting Eq.~(\ref{eq:QH1}) yields the maximum QFI
\begin{equation}\label{eq:result1}
\FImax_\varepsilon = 16\normspec{\coupling^\dagger \GFs^\dagger(\ev) \Hp\GFs(\ev) \coupling}^2\,\intensity .
\end{equation}
This maximum can be achieved by wavefront shaping, i.e., by optimizing the incoming state $\ket{\incoming}$ for fixed~$\intensity$~\cite{BRM21}. For a given scattering system, this maximum QFI in Eq.~(\ref{eq:result1}) is determined by the Green's function~$\GFs$, the perturbation~$\Hp$, and the coupling matrix~$\coupling$, allowing the following physical interpretation: $\coupling$ connects the incoming channels to the system's boundary, the Green's function~$\GFs$ describes the propagation inside the system to the information source $\Hp$ and back to the system's boundary, where $\coupling^\dagger$ connects to the outgoing channels.

It should be noted that this scattering formalism is quite general. In contrast, previous approaches have been limited to very specific model systems~\cite{CKW26,ZSH19}.

\subsection{Localized information source}
In the following, we consider localized perturbations,
\begin{equation}\label{eq:Hplocalized}
\Hp = \ket{j}\bra{j}
\end{equation}
where $\ket{j}$ is a unit vector from a spatial basis $\{\ket{j}\}$ with $j = 1,\ldots N$. In a photonic setup, $j$ may label single-mode cavities. Inserting the localized perturbation~(\ref{eq:Hplocalized}) into Eq.~(\ref{eq:QH1}) gives $\WS_\varepsilon = -\ket{\Qev}\bra{\Qev}$ with the vector
\begin{equation}
\ket{\Qev} := \sqrt{2}\coupling^\dagger\GFs^\dagger\ket{j} .
\end{equation}
The squared norm of this vector is
\begin{equation}\label{eq:psinorm}
	\braket{\Qev}{\Qev} = 2\bra{j}\GFs\coupling\coupling^\dagger\GFs^\dagger\ket{j} .
\end{equation}
Similar to the derivation of Eq.~(\ref{eq:Gdown1}), we derive
\begin{equation}\label{eq:Gdown2}
	\GFs^\dagger-\GFs = 2i\GFs\coupling\coupling^\dagger\GFs^\dagger 
\end{equation}
by multiplying the adjoint of
\begin{equation}\label{eq:GFEH}
\ident_N = \GFs(\ev) (\ev\ident_N-\Hs)
\end{equation}
from the left by $\GFs$ and subtracting the resulting equation from  its adjoint.  Again, one exploits the Hermiticity of $\Hsystem$. We consider the local density of states (LDOS), see e.g. Ref.~\cite{GCM96}, of the unperturbed system 
\begin{equation}\label{eq:LDOSdef}
\DOS_j(\ev) = -\frac{1}{\pi}\imagc \bra{j}\GFs\ket{j} = \frac{i}{2\pi}\bra{j}\GFs-\GFs^\dagger\ket{j} .
\end{equation}
Plugging Eq.~(\ref{eq:Gdown2}) into Eq.~(\ref{eq:psinorm}) and using the LDOS yields
\begin{equation}\label{eq:psinorm2}
	\braket{\Qev}{\Qev} = 2\pi\DOS_j(\ev) .
\end{equation}
With $\WS^2_\varepsilon = \ket{\Qev}\braket{\Qev}{\Qev}\bra{\Qev}$, Eq.~(\ref{eq:FIdefQ}) gives~\cite{HBK21}
\begin{equation}\label{eq:FIuinrho}
	\FI_\varepsilon(\incoming) = 16\pi\DOS_j(\ev) |\bra{j} \GFs\coupling\ket{\incoming}|^2/(\hbar\omega) .
\end{equation}
Note that $\ket{\Qev}$ is an eigenvector of $\WS_\varepsilon$ and therefore also of $\WS^2_\varepsilon$. It is the only one with nonzero eigenvalue. 

For the maximum QFI achievable by wavefront shaping we use Eq.~(\ref{eq:FIuin}) and the fact that $\normspec{\ket{\Qev}\bra{\Qev}}^2 = \braket{\Qev}{\Qev}^2$ yielding 
\begin{equation}\label{eq:FIrho}
\FImax_\varepsilon = 16\pi^2\DOS^2_j(\ev)\intensity .
\end{equation}
Alternatively, one could consider the QFI averaged over all incoming channels, $\FIavg_\varepsilon := \frac{1}{M}\sum_i\FI_\varepsilon(u_i^{\text{in}})$. Using Eqs.~(\ref{eq:Gdown1}) and~(\ref{eq:FIuinrho}), it is straightforward to show that $\FIavg_\varepsilon = \FImax_\varepsilon/M$~\cite{HBK21}. 
We come to the conclusion that the maximum QFI per incoming photon flux, $\FImax_\varepsilon/\intensity$, is directly determined by the LDOS, which is also consistent with Ref.~\cite{BCM20}. As the LDOS can be enhanced at an EP if compared to the corresponding DP~\cite{LPL16,PZM17}, we expect \HL{an enhanced maximal QFI under otherwise comparable conditions.} 

One physical interpretation of the maximum QFI is also that $\sqrt{\FImax_\varepsilon}$ is proportional to  the local delay time defined by $\tau_j(\ev) := 2\pi\DOS_j(\ev)$. Therefore, the longer the light interacts with the information source, the more accurately the parameter $\varepsilon$ can be estimated, as indicated by the Cram\'er-Rao bound~(\ref{eq:CramerRao}).
This interpretation is equivalent to the picture that an incoming channel excites the information source, where Fisher information is created and radiated outwards in terms of a spatial Fisher information flow~\cite{HRR24}.

\section{Fisher information for isolated modes, diabolic, and exceptional points}
\label{sec:FIDPEP}
\subsection{Isolated modes and diabolic points}
\label{sec:FIDP}
Using the self-adjointness, $\normspec{\op{A}^\dagger} = \normspec{\op{A}}$, and the submultiplicativity, $\normspec{\op{A}\op{B}} \leq \normspec{\op{A}}\,\normspec{\op{B}}$, of the spectral norm~\cite{HJ13}, we can derive an upper bound for the maximum QFI in Eq.~(\ref{eq:result1})
\begin{equation}\label{eq:result1b}
\FImax_\varepsilon \leq 16\normspec{\coupling}^4\,\normspec{\GFs(\ev)}^4\, \normspec{\Hp}^2\,\intensity .
\end{equation}
The right hand side is in principle a strict bound, i.e., the equality sign 
\HL{can in principle be achieved for suitably optimized couplings and probing states} by optimizing the matrices $\coupling$ and~$\Hp$ for a given matrix $\GFs$. However, the dependence of $\GFs$ on $\coupling$ limits the possibilities.  

We discuss first the cases of isolated modes and diabolic points. In both cases, the Green's function according to Eq.~(\ref{eq:GKato}) is
\begin{equation}\label{eq:GFDP}
	\GFs(\ev) = \frac{\projector_\ind}{\ev-\evDP} .
\end{equation}
$\evDP$ is the eigenfrequency of the DP, which is $\order$-fold degenerate, or the isolated eigenstate (in which case we may denote the eigenfrequency by $\omega_\ind$). The operator~$\projector_\ind$ projects  onto the generalized eigenspace. In the case of a DP, this eigenspace has dimension $\order$; for an isolated eigenstate, it is one-dimensional.
We introduce the spectral response strength of the isolated mode, the so-called Petermann factor~\cite{Wiersig23b} via 
\begin{equation}\label{eq:defPF}
\PF_\ind = \normspec{\projector_\ind}^2 \ .
\end{equation}
It equals $1$ if $\projector_\ind$ is an orthogonal projector, and is greater than $1$ for the more general case of a non-orthogonal projector. Equivalently, the Petermann factor can be expressed by the more known formula using the right eigenvector $\ket{R_\ind}$ and the left eigenvector $\bra{L_\ind}$ of the unperturbed Hamiltonian
\begin{equation}\label{eq:PF}
\PF_\ind = \frac{1}{|\braket{L_\ind}{R_\ind}|^2} .
\end{equation}
Here, both types of eigenvectors are individually normalized to have a magnitude of one.
With this definition, we plug Eq.~(\ref{eq:GFDP}) into Eq.~(\ref{eq:result1b}) and write, again using the spectral norm, 
\begin{equation}\label{eq:FIDP} 
\FImax_\varepsilon \leq \frac{16}{|\ev-\evDP|^{4}}\normspec{\Hp}^2\normspec{\coupling}^4\PF_\ind^2\,\intensity .
\end{equation}

It should be noted that an increase in the Petermann factor leads to an increase of the upper bound of the QFI, and consequently, in sensitivity. This is counterintuitive at first glance, given that in laser systems there is a linewidth broadening with increasing Petermann factor. However, this does not apply to our case due to the lack of optical gain. What does apply here, is the fact that the Petermann factor is also a measure of eigenvalue sensitivity with respect to perturbations, see e.g.~Refs.~\cite{Wiersig23,KWS25}. 

A more precise and simpler bound can be derived for the case of a localized information source in Eq.~(\ref{eq:Hplocalized}). Inserting the Green's function in Eq.~(\ref{eq:GFDP}) into the LDOS in Eq.~(\ref{eq:LDOSdef}) gives
\begin{equation}
\DOS_j(\ev) = -\frac{1}{\pi}\image{\frac{\bra{j}\projector_\ind\ket{j}}{\ev-\evDP}} .
\end{equation}
An upper bound is
\begin{equation}\label{eq:FIDPlocalized0} 
	\DOS_j(\ev) \leq \frac{1}{\pi}\frac{|\bra{j}\projector_\ind\ket{j}|}{|\ev-\evDP|}\leq \frac{1}{\pi}\frac{\normspec{\projector_\ind}}{|\ev-\evDP|} . 
\end{equation}
With the Petermann factor in Eq.~(\ref{eq:defPF}), the upper bound for the maximal QFI in Eq.~(\ref{eq:FIrho}) is
\begin{equation}\label{eq:FIDPlocalized} 
\FImax_\varepsilon \leq 16 \frac{\PF_\ind}{|\ev-\evDP|^2}\,\intensity .
\end{equation}
In addition to the frequency detuning $|\ev - \realc{\evDP}|$, there are here two further independent knobs for controlling the QFI: the decay rate~$|\imagc{\evDP}|$ and the strength of the spectral response expressed by the Petermann factor $\PF_\ind$. The adjustment of information source and scattering states no longer appears in Eq.~(\ref{eq:FIDPlocalized}). It is eliminated in the final step in Eq.~(\ref{eq:FIDPlocalized0}), where the LDOS is bounded by a term that is independent  of the spatial basis vector $\ket{j}$.
The bound~(\ref{eq:FIDPlocalized}) for localized information sources is considerably simpler than the more general one in inequality~(\ref{eq:FIDP}) and, as we see later, more precise.

\subsection{Exceptional points}
Next, we specify the unperturbed system to have an EP of order~$\order$. We assume that it is sufficient to consider the Green's function near the EP in Eq.~(\ref{eq:GatEP}). In the present setup, the frequency $\omega$ is real-valued, hence for passive systems (no gain), where $\evEP$ might be deep in the complex plane, this condition may not always be achievable. 
\HL{Here, this is not a significant limitation, since a high-performance sensor in this context is expected to exhibit a small decay rate. Moreover, the assumption also relies on the requirement $\coupling^\dagger (\nilpotent^\dagger)^{\order-1} \Hp\nilpotent^{\order-1} \coupling \neq 0$, see Eq.~(\ref{eq:result1}), i.e. the scattering states and information source should be properly adjusted.}

With this assumption and the definition of the spectral response strength in Eq.~(\ref{eq:rcagen}) we get from Eq.~(\ref{eq:result1b})  for the resonant case the upper bound
\begin{equation}\label{eq:FIEP} 
\FImax_\varepsilon \leq \frac{16}{|\ev-\evEP|^{4\order}}\normspec{\Hp}^2\,\normspec{\coupling}^4\,\rca^4\,\intensity .
\end{equation}
Also here, we consider the special case of a localized information source in Eq.~(\ref{eq:Hplocalized}). Inserting the Green's function in Eq.~(\ref{eq:GatEP}) into the LDOS in Eq.~(\ref{eq:LDOSdef}) gives
\begin{equation}
	\DOS_j(\ev) = -\frac{1}{\pi}\image{\frac{\bra{j}\nilpotent^{\order-1}\ket{j}}{(\ev-\evEP)^\order}} .
\end{equation}
An upper bound is
\begin{equation}
	\DOS_j(\ev) \leq \frac{1}{\pi}\frac{|\bra{j}\nilpotent^{\order-1}\ket{j}|}{|\ev-\evEP|^\order}\leq \frac{1}{\pi}\frac{\normspec{\nilpotent^{\order-1}}}{|\ev-\evEP|^\order} . 
\end{equation}
With the spectral response strength in Eq.~(\ref{eq:rcagen}), the upper bound for the QFI in Eq.~(\ref{eq:FIrho}) is
\begin{equation}\label{eq:FIEPlocalized} 
\FImax_\varepsilon \leq 16 \frac{\rca^2}{|\ev-\evEP|^{2\order}}\,\intensity .
\end{equation}
Again, on resonance $\ev = \realc{\evEP}$ this bound depends on the decay rate~$|\imagc{\evEP}|$ and the spectral response strength, here~$\rca$. 
Comparing the bound~(\ref{eq:FIEPlocalized}) with the one in the DP case in Eq.~(\ref{eq:FIDPlocalized}) with $\evEP = \evDP$ for the resonant case $\ev = \realc{\evEP}$ we get for the most extreme case $\PF = 1$ the enhancement factor for the upper bounds
\begin{equation}\label{eq:EF}
\text{EF} = \frac{\rca^2}{|\imagc{\evEP}|^{2(\order-1)}} .
\end{equation}

\subsection{Passive systems}
For a Hamiltonian of a passive system there is the following restriction for the spectral response strength~\cite{Wiersig22b}
\begin{equation}\label{eq:rcapassve}
	\rca \leq \left(\sqrt{2n}|\imagc{\evEP}|\right)^{\order-1} .
\end{equation}
This restriction is only valid for a system whose Hilbert-space dimension equals the order of the EP, i.e., beside the EP there are no other states~\cite{Wiersig23b}.
The upper bound in Eq.~(\ref{eq:rcapassve}) is not strict for $\order > 2$. For $\order = 3$ a strict upper bound has been determined numerically~\cite{Wiersig22}
\begin{equation}\label{eq:rcapassve3}
\rca \leq 4|\imagc{\evEP}|^2 .
\end{equation}
Here, and for the case $\order = 2$ in inequality~(\ref{eq:rcapassve}) the equal sign holds when the decay operator $\oHami := i(\op{H}-\op{H}^\dagger)$ is a rank-1 matrix. In this case the maximum spectral response strength for a given decay rate is attained. 
 
The condition~(\ref{eq:rcapassve}) carries over to the enhancement factor in Eq.~(\ref{eq:EF})
\begin{equation}\label{eq:EFpassive}
	\text{EF} \leq (2\order)^{\order-1} ,
\end{equation}
which is $4$ for a second-order EP and $36$ [$16$ if using Eq.~(\ref{eq:rcapassve3})] for a third-order EP.
Note that, in practice, the QFI may be smaller than the upper bounds derived above. Therefore, comparing the QFI in the EP and non-EP cases may not yield an enhancement factor exactly as given by Eq.~(\ref{eq:EFpassive}). Nevertheless, Eq.~(\ref{eq:EFpassive}) gives a clear indication that even for passive systems a significant enhancement of the maximum QFI at an EP, if compared to the corresponding DP, is possible. 

\section{Examples: Microring-waveguide systems}
\label{sec:examples}
We now apply the general formalism developed above to specific microring-waveguide sensing platforms. These systems provide experimentally relevant realizations of isolated modes, diabolic points, and exceptional points, and thus offer a concrete setting in which to illustrate the bounds and scaling behavior derived in the previous sections.

\subsection{Two coupled microrings}
\label{sec:microringEP2}
\begin{figure}[ht]
 \centering\includegraphics[width=0.5\columnwidth]{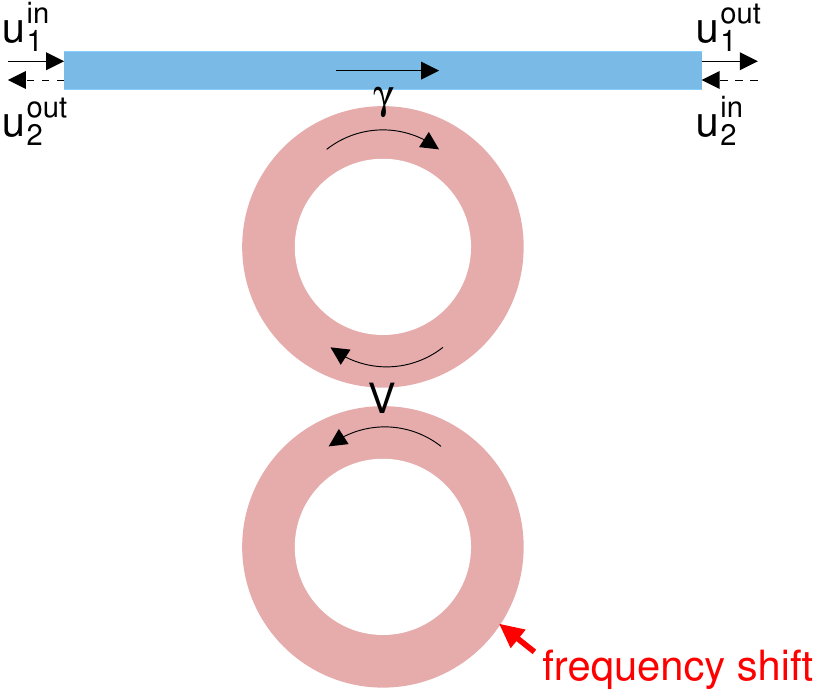}
\caption{Sketch of two coupled optical microrings attached to a single-mode waveguide with coupling strength $\gamma$. $V$ is the intercavity coupling coefficient. The information source is located in the lower ring, i.e., the system is perturbed by shifting its frequency with respect to the upper ring, for instance by a micro-heater underneath the lower ring, see Refs.~\cite{HHW17,FAB25}.} 
\label{fig:examplecoupledring}
\end{figure}

The possibly simplest nontrivial example is the setup of two coupled microrings illustrated in Fig.~\ref{fig:examplecoupledring}. 
An incoming state from the left along a single-mode waveguide couples into the clockwise (CW) propagation direction of the upper ring. This wave then couples into the counterclockwise (CCW) propagation in the lower ring. Subsequently, this wave couples back into the CW direction of the upper ring, which in turn couples out to the waveguide propagating to the right. An incoming state from the right exhibits a similar behavior, but with the CW and CCW directions exchanged. As a result, the four-mode system decouples into two separate two-mode systems. Alternatively, one could also start directly from two coupled single-mode cavities as has been done in Refs.~\cite{CJL19,LS24,DMA22}.

For the first two-mode system, the matrix describing the coupling between incoming channels and microring is
\begin{equation}
	\coupling =  \sqrt{\frac{\gamma}{2}} \left(\begin{array}{c}
		1 \\
		0 \\
	\end{array}\right) ,
\end{equation}
where $\gamma > 0$ is the strength of the coupling between waveguide and upper microring.
Assuming microrings with identical frequencies $\omega_0 \in\R$ the unperturbed Hamiltonian [Eq.~(\ref{eq:Heff})] is 
\begin{equation}\label{eq:Hscr}
	\Hs =  \left(\begin{array}{cc}
		\omega_0-i\frac{\gamma}{2} & V^* \\
		V & \omega_0 \\
	\end{array}\right) ,
\end{equation}
where $V \in\C$ is the intercavity coupling coefficient, and $|V|$ its strength. Here, we assume that material absorption and radiation losses can be ignored as the corresponding decay rate is much smaller than $\gamma$. It is important to understand that the non-Hermiticity of $\Hs$ originates from the coupling to scattering channels described by the $-i\coupling\coupling^\dagger$-term in Eq.~(\ref{eq:Heff}). As a result, the scattering matrix is unitary.
  
The eigenvalues of the unperturbed Hamiltonian~(\ref{eq:Hscr}) are
\begin{equation}\label{eq:evH}
\ev_\pm = \omega_0-i\frac{\gamma}{4}\pm\sqrt{|V|^2-\frac{\gamma^2}{16}} .
\end{equation}
For $|V| = \gamma/4$ these two eigenvalues degenerate to $\evEP = \omega_0 - i\frac{\gamma}{4}$ at an EP of order $n=2$. The spectral response strength is $\rca = 2|V| = \gamma/2$~\cite{Wiersig22}. This is the maximal strength achievable for the given decay rate $\gamma/4$ since the decay operator $\oHami = 2\coupling\coupling^\dagger$ is here of rank 1. 

Similar systems of two coupled single-mode cavities were analyzed in Refs.~\cite{CJL19,LS24,DMA22,AAK25}, though with additional cavity losses (except~\cite{AAK25}) – a case we will examine in a later section. We consider as perturbation a frequency shift of the traveling waves in the lower ring, e.g. by local heating~\cite{HHW17,FAB25}, described by the Hermitian operator
\begin{equation}\label{eq:HPtworing}
\varepsilon\Hp =  \left(\begin{array}{cc}
0 & 0\\
0 & \varepsilon \\
\end{array}\right) .
\end{equation}
It turns out that this perturbation, in the present setup, gives the largest QFI. The same holds true for a symmetric detuning of the frequencies of both rings, as considered in Ref.~\cite{CJL19}. A frequency perturbation applied solely to the first ring or perturbing the intercavity coupling, i.e. varying $V$, as considered in Refs.~\cite{LS24,DMA22,AAK25}, is not the best choice in our context. This is because, in the lossless limit considered here, it would result in a vanishing QFI. The mathematical reason is $\Hp\GFs(\evEP)\coupling = 0$ in Eq.~(\ref{eq:result1}) or, for a localized information source, a vanishing LDOS in Eq.~(\ref{eq:FIrho}).
\HL{This demonstrates that the adjustment between the scattering states and the information source is crucial. From a practical standpoint, it implies that a localized information source should be positioned at the point where the LDOS is maximal.}

\begin{figure}[ht]
\centering\includegraphics[width=0.75\columnwidth]{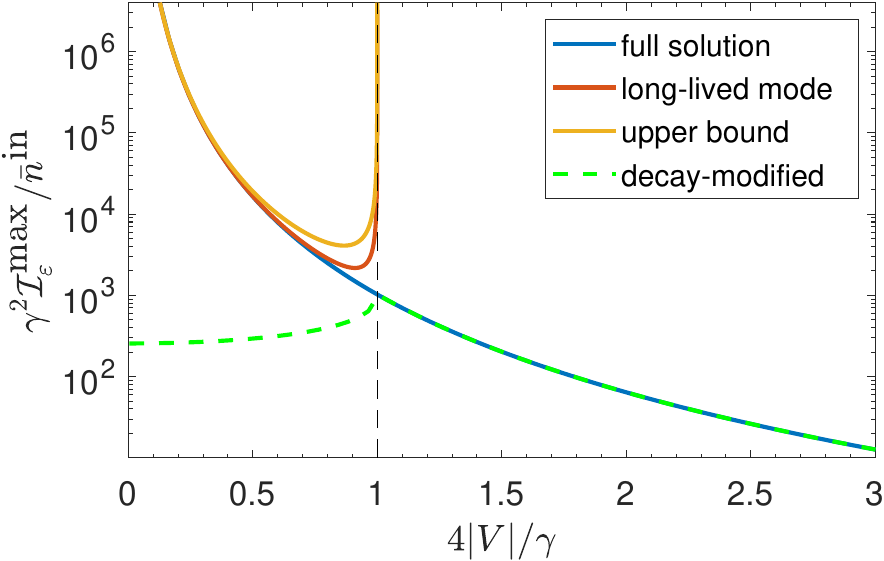}
\caption{The maximal QFI for the two-ring setup as function of the intercavity coupling strength $|V|$ on a semi-logarithmic scale. Both quantities are made dimensionless by scaling with the waveguide coupling strength~$\gamma$. The blue curve results from the full calculation according to Eq.~(\ref{eq:ecr0}), while the red curve shows the contribution from the long-lived mode [one summand in Eq.~(\ref{eq:LDOSRL})] in the weak coupling regime. The orange curve is the upper bound of this contribution, as given by Eq.~(\ref{eq:FIDPlocalized}). The green dashed curve shows the decay-modified QFI from Eq.~(\ref{eq:FImod}). The EP is indicated by the dashed vertical line.}
\label{fig:IvsV}
\end{figure}

Inserting the Green's function $\GFs$ [based on $\Hs$ via Eq.~(\ref{eq:G0})], $\coupling$, $\Hp$, and $\omega = \omega_0$ into Eq.~(\ref{eq:result1}) or~(\ref{eq:FIrho}) gives
\begin{equation}\label{eq:ecr0}
	\FImax_\varepsilon = 4\frac{\gamma^2}{|V|^4}\,\intensity .
\end{equation}
Hence, the QFI decays monotonically with the coupling strength $|V|$, which is shown in Fig.~\ref{fig:IvsV}.  Nothing particular seems to be happening at the EP $|V| = \gamma/4$ for which we get from Eq.~(\ref{eq:ecr0})
\begin{equation}\label{eq:ecr1}
	\FImax_\varepsilon = \frac{1024}{\gamma^2}\,\intensity \quad\text{(EP)}.
\end{equation}

Based on a missing peak at the EP, Ref.~\cite{AAK25} (computing and measuring a quantity which is similar to a classical version of the QFI) concluded that the sensing performance is not enhanced at the EP. However, this raises the crucial question: enhanced compared to what? The original proposal of EP-based sensing~\cite{Wiersig14b} highlights the importance of comparing the EP to a DP with identical eigenfrequencies. In our case, there is no DP for $\gamma \neq 0$. 
Therefore, we compare to an isolated mode that shares the same characteristics as the DP discussed in Sec.~\ref{sec:FIDP}. In particular, it is described by the same form of the Green's function~(\ref{eq:GFDP}), which implies a linear scaling of the eigenfrequency(ies) under small perturbations. For simplicity, we consider a single-ring setup with the information source being a frequency shift of the mode in the only ring left. For a fair comparison, the coupling to the waveguide is reduced by a factor of two, such that $|\imagc{\ev_1}| = \gamma/4 = |\imagc{\evEP}|$. Note that this adjustment does not favor the EP. Using Eq.~(\ref{eq:result1}) yields
\begin{equation}\label{eq:ecrsingle}
\FImax_\varepsilon = \frac{256}{\gamma^2}\,\intensity \quad\text{(isolated mode)}.
\end{equation}
Comparing to Eq.~(\ref{eq:ecr1}) reveals that the QFI is enhanced at the EP by a factor of 4 in agreement with the general estimation in Eq.~(\ref{eq:EFpassive}).
\HL{Note that this comparison at a fixed decay rate is intended solely to disentangle the distinct mechanisms underlying the QFI enhancement, namely, those arising from non-normality and those associated with lifetime effects.}

Using Eq.~(\ref{eq:FIEP}) gives the general upper bound for the EP in the two-ring setup
\begin{equation}\label{eq:ecrxi}
	\FImax_\varepsilon \leq \frac{16384}{\gamma^2}\,\intensity .
\end{equation}
It can be verified that the inaccuracy of this upper bound arises from utilizing the submultiplicativity property of the matrix norm. The bound is not very tight, as it is 16 times larger than the actual value.
Using the upper bound for a localized information source in Eq.~(\ref{eq:FIEPlocalized}) gives
\begin{equation}\label{eq:ecrxilocalized}
	\FImax_\varepsilon \leq \frac{1024}{\gamma^2}\,\intensity ,
\end{equation}
which perfectly agrees with the exact value in Eq.~(\ref{eq:ecr1}).

One crucial question remains: What is the physical mechanism that allows for a further increase of the QFI in Eq.~(\ref{eq:ecr0}) by reducing the intercavity coupling strength~$|V|$ beyond the EP? Starting at the EP with $|V| = \gamma/4$, the system enters the weak coupling regime $|V| < \gamma/4$ when the parameter~$|V|$ is reduced, cf. Eq.~(\ref{eq:evH}), resulting in a linewidth splitting. One eigenvalue moves deeper into the complex frequency plane, while the other one moves closer to the real axis until it finally reaches it. This scenario, approaching the case of a bound state in the continuum (BIC)~\cite{FriWin85,Wiersig06}, gives rise to an increasingly narrow resonant phase shift in transmission that is very sensitive with respect to any system perturbation and thus produces much Fisher information at the output. Note that the limit $\gamma\to 0$ is singular, as for $\gamma = 0$ the incoming and outgoing scattering states are completely decoupled from the information source.
The movement of the eigenfrequency towards the real axis increases the LDOS on resonance, which according to Eq.~(\ref{eq:FIrho}) increases the QFI. 
Although the system is not exactly at the EP in this case, the linewidth splitting is a result of its proximity to an EP; a similar phenomenon can be observed near the laser threshold as discussed in Refs.~\cite{RotterNC14,KZW25}.

For an alternative perspective, we express the Green's function~(\ref{eq:GKato}) of the unperturbed Hamiltonian, not at the EP, in terms of normalized right eigenvectors $\ket{R_\ind}$ and left eigenvectors $\bra{L_\ind}$ 
\begin{equation}\label{eq:GnotEP}
	\GFs(\ev) = \sum_\ind\frac{\ket{R_\ind}\bra{L_\ind}}{\braket{L_\ind}{R_\ind}} \frac{1}{\ev-\ev_\ind}.
\end{equation}
Inserting this for the central frequency~$\omega_0$ into the LDOS in Eq.~(\ref{eq:LDOSdef}) yields
\begin{equation}\label{eq:LDOSRL}
\DOS_j(\omega_0) = -\frac{1}{\pi}\imagc \sum_\ind\frac{\braket{j}{R_\ind}\braket{L_\ind}{j}}{\braket{L_\ind}{R_\ind}} \frac{1}{\omega_0-\ev_\ind} .
\end{equation}
Importantly, each summand in Eq.~(\ref{eq:LDOSRL}) diverges to infinity when approaching the EP due to the selforthogonality $\braket{L_\ind}{R_\ind} = 0$ at the EP~\cite{SAC06}. Equally, this can be expressed by the diverging Petermann factors~(\ref{eq:PF}). Considering this divergent behavior, one would expect the EP to always provide the largest QFI; however, this is not what is observed in Eq.~(\ref{eq:ecr0}). This apparent contradiction is resolved by recognizing that such infinities must cancel in the sum, as the LDOS at the EP, while large, also remains finite as long as no eigenfrequency is located on the real axis~\cite{YSS11}; see also Refs.~\cite{Wiersig23b,KWS25}. Even a slight departure from the EP can disturb this fragile balance of divergences in Eq.~(\ref{eq:LDOSRL}), potentially leading to an increase in the LDOS and therefore also the QFI, despite the reduction in the -- now finite -- Petermann factors.   

To appreciate this line of reasoning, consider the weak coupling regime. Here, the balance of divergences is disrupted as the modes become unevenly distributed across the two cavities, with the longer-lived mode becoming more localized in the lower ring. The corresponding summand in Eq.~(\ref{eq:LDOSRL}) will dominate as its eigenfrequency comes closer to the real axis. Hence, the QFI is enhanced by linewidth narrowing resulting from linewidth splitting.
This scenario is supported by the red curve in Fig.~\ref{fig:IvsV}, which shows the contribution of the long-lived mode to the QFI~(\ref{eq:FIrho}) through the LDOS~(\ref{eq:LDOSRL}). 
Far from the EP (in the limit $|V| \to 0$), this contribution dominates the QFI. However, as we approach the EP, the curve diverges to infinity. As a result, the shorter-lived mode must compensate for this divergence. This is the balance of divergences discussed above. The orange curve demonstrates that Eq.~(\ref{eq:FIDPlocalized}) provides a reliable upper bound for the QFI when a single isolated mode is dominant.

Next, in Fig.~\ref{fig:IvsV}, we plot an artificial quantity that we refer to as the decay-modified QFI. To motivate this definition, consider first the result for the isolated mode in Eq.~(\ref{eq:ecrsingle}), which shows that the QFI increases as the decay rate of the mode, $\gamma/4$.  This behavior arises simply because a narrower linewidth improves the precision of the sensing. To remove this increase of the QFI caused solely by linewidth narrowing, we could multiply $\FImax_\varepsilon$ in the isolated-mode case by $\gamma^2/16$. The resulting quantity is not directly measurable in experiments; it is introduced in the following for the two-ring setup solely to separate the effects of non-Hermiticity (decay rate) from those of non-normality (spectral response strength). Here, we define the decay-modified QFI as
\begin{equation}\label{eq:FImod}
 \FImod_\varepsilon := \FImax_\varepsilon 16|\imagc\ev_+|^2
\end{equation}
with $\FImax_\varepsilon$ from Eq.~(\ref{eq:ecr0}) and the smallest decay rate $|\imagc\ev_-|$ according to Eq.~(\ref{eq:evH}). The factor of $16$ is chosen to facilitate a clearer comparison with $\gamma^2\FImax$ in  Fig.~\ref{fig:IvsV}. In the strong coupling regime, $|V| > \gamma/4$, the decay-modified QFI coincides with the maximum QFI because the decay rate, $|\imagc\ev_-| = \gamma/4$, is independent of $|V|$. In contrast, in the weak coupling regime, $|V| < \gamma/4$, the decay rate $|\imagc\ev_-|$ decreases due to the linewidth splitting as $|V|$ decreases. This reduction is sufficient to lower the decay-modified QFI defined in Eq.~(\ref{eq:FImod}). As a result, the decay-modified QFI exhibits a maximum at the EP. This demonstrates that in the present setup the experimentally accessible QFI, $\FImax_\varepsilon$, does not exhibit a maximum exactly at the EP since moving away from the EP can further enhance the QFI through linewidth narrowing.

\HL{
Let us emphasize that the decay-modified QFI as well as the comparison to a DP are merely diagnostic tools used to elucidate the underlying physics of non-Hermitian sensors. For practical sensor design, the central objective is not to realize artificially matched decay rates or to operate precisely at an EP, but rather to maximize the achievable QFI within the relevant experimental constraints.
}

\begin{figure}[ht]
\centering\includegraphics[width=0.75\columnwidth]{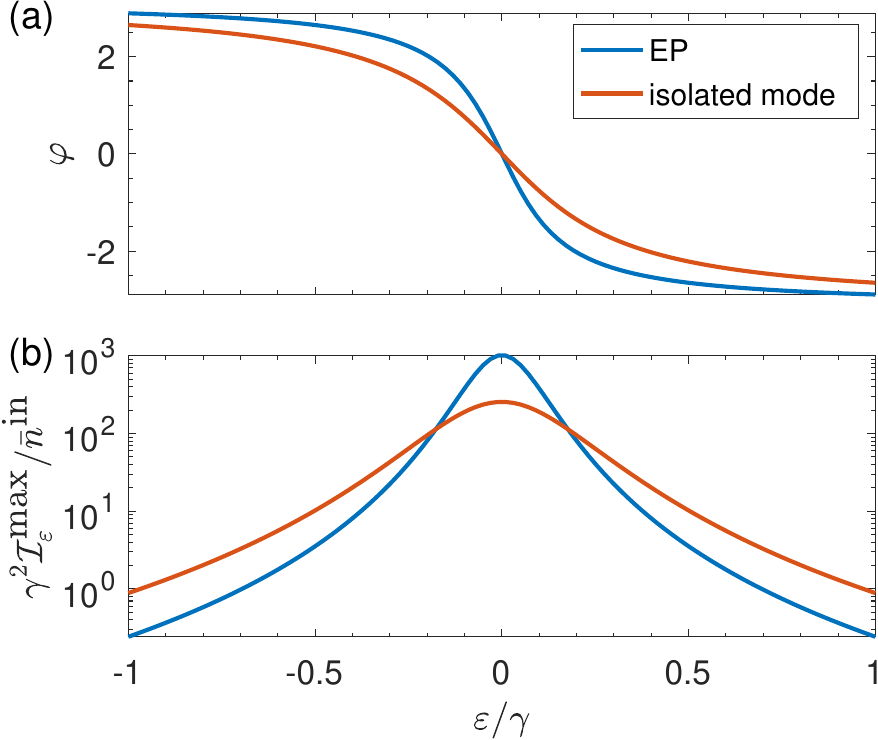}
\caption{(a) The phase $\varphi$ in radians of the outgoing state as function of the normalized perturbation parameter $\varepsilon$, which equals a detuning of the lowermost ring for the EP in the two-ring setup illustrated in Fig.~\ref{fig:examplecoupledring} and in the single-ring setup with the isolated mode.
(b) Maximal QFI in Eq.~(\ref{eq:ecrphase}) made dimensionless by scaling with the waveguide coupling strength~$\gamma$.
Without loss of generality, the phase $\varphi$ is gauged such that it is zero at $\varepsilon = 0$. The frequency is chosen to be on resonance, $\omega = \omega_0$.}
\label{fig:phase}
\end{figure}

In the case of a single incoming and outgoing channel, as we have here because CW and CCW propagation directions separate, the unitary scattering matrix is a complex number of modulus one, i.e., $\Smatrix = e^{i\varphi}$ with the phase~$\varphi\in [-\pi,\pi)$. Mathematically, this is obvious but the physical implication is surprising as it means that there is no dip in the intensity transmission spectrum even though the light couples resonantly into and out of the microrings. The physical reason is that there are no internal losses in the microrings and no back reflections to the input port. Experimentally, only a phase change occurs that can be measured by mapping it to an intensity change via an interferometric setup.
We can write the QFI in Eq.~(\ref{eq:FIdef}) as
\begin{equation}\label{eq:ecrphase}
\FImax_\varepsilon = 4(\partial_\varepsilon\varphi)^2\,\intensity .
\end{equation}
Figure~\ref{fig:phase}(a) shows the phase~$\varphi$ both for the EP and the isolated mode, plotted as a function of the perturbation parameter~$\varepsilon$. The slope at the position of interest, $\varepsilon = 0$, is larger for the EP which translates, according to Eq.~(\ref{eq:ecrphase}), to a larger QFI as depicted in Fig.~\ref{fig:phase}(b). 
Note that in this parameter space, the QFI exhibits a clear maximum at the EP. It decays when leaving the EP by increasing $|\varepsilon|$.

\subsection{Three coupled microrings}
\label{sec:microringEP3}
In the same spirit, we consider now three coupled microrings (not illustrated) with coupling matrix 
\begin{equation}
	\coupling =  \sqrt{\frac{\gamma}{2}} \left(\begin{array}{c}
		1 \\
		0 \\
		0 \\
	\end{array}\right) ,
\end{equation}
where again $\gamma > 0$ is the strength of the coupling between waveguide and uppermost microring. Assuming microrings with identical frequencies $\omega_0\in\R$  the unperturbed Hamiltonian [Eq.~(\ref{eq:Heff})] is 
\begin{equation}\label{eq:Hs3cr}
\Hs =  \left(\begin{array}{ccc}
\omega_0-i\frac{\gamma}{2} & V_1^* & 0 \\
V_1 & \omega_0 & V_2^*\\
0 & V_2 & \omega_0\\
	\end{array}\right) ,
\end{equation}
where $V_1, V_2 \in\C$ are the intercavity coupling coefficients. Again, the non-Hermiticity of $\Hs$ originates only from the coupling to the scattering channels.

The Hamiltonian~(\ref{eq:Hs3cr}) possesses a third-order EP for $V_1 = \frac{\sqrt{2}\gamma}{3\sqrt{3}}$ and $V_2 = \frac{\gamma}{6\sqrt{3}}$~\cite{WLX24}. The eigenvalue at the EP is $\evEP = \omega_0 - i\frac{\gamma}{6}$ and, with the methods in Ref.~\cite{Wiersig22}, the spectral response strength can be calculated to be $\rca = \gamma^2/9$. Again, for given decay rate, $\gamma/6$, this is the largest possible spectral response strength as the decay operator $\oHami$ is of rank 1. 

As the perturbation, we once again consider a frequency shift of the lowest ring described by the Hermitian operator
\begin{equation}
\varepsilon\Hp =  \left(\begin{array}{ccc}
0 & 0 & 0\\
0 & 0 & 0\\
0 & 0 & \varepsilon \\
\end{array}\right) .
\end{equation}
This turns out to be the best choice, aside from scenarios where all rings are perturbed simultaneously, which may be challenging to achieve in practice.

Inserting $\coupling$, $\Hs$, $\Hp$ and $\omega = \omega_0$ into Eq.~(\ref{eq:result1}) gives
\begin{equation}\label{eq:ecr2}
\FImax_\varepsilon = \frac{4096}{\gamma^2}\,\intensity \quad\text{(EP)} .
\end{equation}
While the general upper bound in Eq.~(\ref{eq:FIEP}) is again rather inaccurate (not shown), the upper bound for the localized information sourcen in Eq.~(\ref{eq:FIEPlocalized}) gives 
\begin{equation}\label{eq:ecrxi3localized}
\FImax_\varepsilon \leq \frac{9216}{\gamma^2}\,\intensity .
\end{equation}

We compare the result of the third-order EP with the isolated mode in a single microring with adjusted waveguide coupling $\gamma/6$ (instead of $\gamma/2$) using Eq.~(\ref{eq:result1}) yielding
\begin{equation}
	\FImax_\varepsilon = \frac{576}{\gamma^2}\,\intensity \quad\text{(isolated mode)} .
\end{equation}
Comparing this result to Eq.~(\ref{eq:ecr2}) reveals that the QFI is enhanced at the EP by a factor of $64/9 \approx 7.1$ consistent with the general estimation in Eq.~(\ref{eq:EFpassive}). Hence, we conclude that a third-order EP in our setup results in an even greater enhancement of the QFI compared to the isolated mode.

\subsection{Fully asymmetric backscattering}
\label{eq:asymback}
\begin{figure}[ht]
\centering\includegraphics[width=0.5\columnwidth]{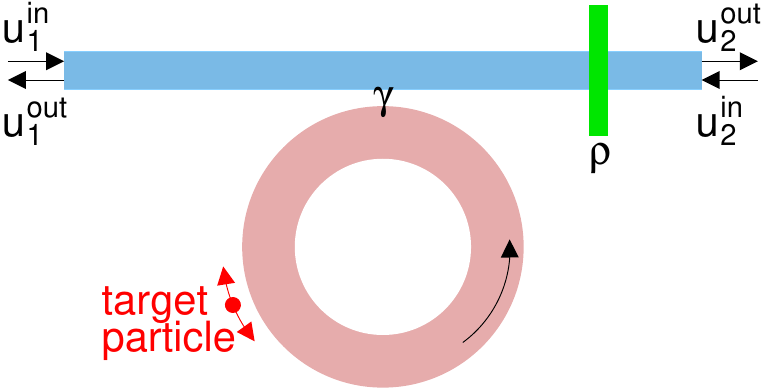}
\caption{Sketch of an optical microring coupled to a semi-infinite single-mode waveguide with a partial mirror having a reflection coefficient~$\rho$; $\gamma$ is the waveguide coupling strength. A small target particle induces a backscattering between clockwise and counterclockwise propagating waves inside the microring.}
\label{fig:exampleWhaley}
\end{figure}

As further example, we study a microring coupled to a semi-infinite waveguide terminated at one site by a partial mirror with reflection coefficient $0\leq \rho \leq 1$, see Fig.~\ref{fig:exampleWhaley}. This geometry results in a so-called exceptional surface, i.e., a surface of EPs in parameter space,  with increased robustness with respect to unwanted perturbations~\cite{ZRK19,SZM22}.

Following the discussion in Ref.~\cite{CKW26} with an adapted notation, the coupling matrix is
\begin{equation}
\coupling =  \sqrt{\frac{\gamma}{2}} \left(\begin{array}{cc}
1               & 0\\
\rho e^{i2\phi} & \tau e^{i\phi}\\
\end{array}\right) ,
\end{equation}
with the traveling-wave basis $\ket{\psi} = (\psi_{\text{CW}},\psi_{\text{CCW}})^\transpose$ for the internal modes of the microring. Again $\gamma > 0$ is the strength of the waveguide coupling, $\phi$ is the phase accumulated by propagating from the microring to the mirror, and $\tau = \sqrt{1-\rho^2}$ is the transmission coefficient through the mirror. 

The unperturbed Hamiltonian including the $-i\coupling\coupling^\dagger$-term is
\begin{equation}\label{eq:HsWhaley}
	\Hs =  \left(\begin{array}{cc}
		\omega_0-i\frac{\gamma}{2} & 0 \\
		-i\gamma\rho e^{i2\phi} & \omega_0-i\frac{\gamma}{2} \\
	\end{array}\right) ,
\end{equation}
where $\omega_0\in\R$ is the frequency of the traveling-wave modes. There is no backscattering from CW to CCW propagation direction. In the special case $\rho = 0$, the unperturbed Hamiltonian in Eq.~(\ref{eq:HsWhaley}) is at an DP with eigenvalue $\evDP = \omega_0 - i\frac{\gamma}{2}$. Increasing $\rho > 0$ turns the DP immediately into an EP of order $n=2$ and eigenvalue $\evEP = \omega_0 - i\frac{\gamma}{2}$~\cite{Wiersig18b}. The spectral response strength of this EP is $\rca = \gamma\rho$~\cite{Wiersig22}. According to Eq.~(\ref{eq:rcapassve}) we have for passive systems $\rca \leq \gamma$, which implies $\rho \leq 1$, consistent with the interpretation of $\rho$ being a reflection coefficient of a gainless mirror. 
Note that the system described by the Hamiltonian~$\Hs$ in Eq.~(\ref{eq:HsWhaley}) in the chosen basis of traveling waves fulfills Lorentz reciprocity, despite being represented by an asymmetric matrix.  In contrast, if a standing-wave basis were employed, the system would be represented by a complex symmetric matrix~\cite{WES11}. 

A small particle near the ring induces backscattering in both propagation directions described by a Hermitian operator~\cite{Wiersig11}
\begin{equation}
	\varepsilon\Hp =  \left(\begin{array}{cc}
		0 & \varepsilon \\
		\varepsilon & 0 \\
	\end{array}\right) .
\end{equation}
This is not a localized information source in the sense of Eq.~(\ref{eq:Hplocalized}), therefore Eq.~(\ref{eq:FIrho}) cannot be used. Instead we insert $\coupling$, $\Hs$, $\Hp$ and $\omega = \omega_0$ into the more general Eq.~(\ref{eq:result1}) yielding the maximal QFI 
\begin{equation}\label{eq:ecrWhaley}
\FImax_\varepsilon = \frac{64(1+\rho)^2}{\gamma^2}\,\intensity .
\end{equation}
As mentioned in Sec.~\ref{sec:FIperturbations} this maximum is achieved by optimizing the incoming state $\ket{\incoming}$ for fixed average photon number~$\intensity$.
In Eq.~(\ref{eq:ecrWhaley}) the dependence on the decay rate $\gamma/2$ and the spectral response strength of the EP, $\rca = \gamma\rho$, is transparent.
The same result as in Eq.~(\ref{eq:ecrWhaley}) was obtained in Ref.~\cite{CKW26} using a customized theoretical framework 
for this particular system. In the limiting case of a reflectionless mirror, $\rho \to 0$, the EP becomes a DP. Increasing $\rho$ increases the spectral response strength of the EP, $\rca = \gamma\rho$. In the limiting case $\rho \to 1$ the maximal spectral response strength is achieved. The QFI in Eq.~(\ref{eq:ecrWhaley}) is then larger by a factor of 4 if compared to the DP case. This enhancement factor is fully consistent with our general result in Eq.~(\ref{eq:EFpassive}).

\section{Internal losses}
\label{sec:losses}
So far, we have focused on the situation where the internal losses -- radiation and absorption losses -- of the cavities are much smaller than the coupling strength~$\gamma$ to the waveguide. However, Eqs.~(\ref{eq:ecr1}), (\ref{eq:ecrsingle}), (\ref{eq:ecr2}), and~(\ref{eq:ecrWhaley}) show that the QFI can be increased by reducing~$\gamma$, which can be done experimentally simply by increasing the distance between waveguide and cavities. Clearly, the divergence of the QFI in the limit $\gamma\to 0$ is not physical. It originates precisely from ignoring the internal losses. In this section, we therefore examine their effect on the QFI. 

Here, we focus on internal losses that can be realistically modeled as additional auxiliary scattering channels. These types of losses do not alter the noise characteristics, allowing our scattering formalism to remain applicable. Radiation losses, as well as certain forms of absorption losses, fall into this category. For simplicity, we restrict ourselves to uniform internal losses where each cavity has its own single decay channel characterized by the same coupling rate $\kappa$. As these additional channels cannot be monitored experimentally, we consider in the following the reduced QFI~\cite{BRM21}, which, moreover, is computed here for light entering only from the left into the waveguide ($\ket{\incoming_1}$) and observing, in the presence of the internal loss channels, only the outgoing light along the waveguide to the right ($\ket{\outgoing_1}$) as e.g. shown in the setup in Fig.~\ref{fig:examplecoupledring}. Using Eqs.~(\ref{eq:FIdef}) and~(\ref{eq:dsm}) adapted to the above situation, we find for the localized information source in Eq.~(\ref{eq:Hplocalized})
\begin{equation}\label{eq:Iredsplit}
\FIred_\varepsilon = 16|\bra{\outgoing_1}\coupling^\dagger\GFs\ket{j}|^2|\bra{j}\GFs\coupling\ket{\incoming_1}|^2 /(\hbar\omega) .
\end{equation}
The factor $|\bra{j}\GFs\coupling\ket{\incoming_1}|^2$ is the intensity $|\psi^{\text{in}}_j|^2$ at the information source resulting from the coupling to the incoming channel 1 in the waveguide. The adjacent factor on the left hand side quantifies the coupling from the information source to the outgoing channel 1 in the waveguide. For the present systems of two and three coupled rings these two factors are proportional to each other due to the spatial symmetry and Lorentz reciprocity. Hence, the reduced QFI directly reflects the behavior of the (square of the) intensity~$|\psi^{\text{in}}_j|^2$ at the information source.

For the two-ring setup we obtain with the unperturbed Hamiltonian 
\begin{equation}\label{eq:Hscrkappa}
	\Hs =  \left(\begin{array}{cc}
		\omega_0-i\frac{\gamma+\kappa}{2} & V^* \\
		V & \omega_0-i\frac{\kappa}{2} \\
	\end{array}\right) ,
\end{equation}
with spatially uniform radiation loss rate $\kappa$, and Eq.~(\ref{eq:Iredsplit})
\begin{equation}\label{eq:ecr1glossesred}
\FIred_\varepsilon = \frac{4|V|^4\gamma^2}{((\gamma+\kappa)\kappa/4+|V|^2)^4}\,\intensity .
\end{equation}
At the EP with $|V| = \gamma/4$ one gets
\begin{equation}\label{eq:ecr1lossesredEP}
\FIred_\varepsilon = 1024\frac{\gamma^6}{(\gamma+2\kappa)^8}\,\intensity \quad\text{(EP)}.
\end{equation}
In the limit of vanishing radiation losses, Eq.~(\ref{eq:ecr1}) is recovered. 

For the single-ring setup with an isolated mode we again chose a waveguide coupling reduced by a factor of 2 if compared to the two-ring setup to ensure a fair comparison. We get for the reduced QFI 
\begin{equation}\label{eq:ecrsinglelossesred}
\FIred_\varepsilon = 16\frac{\gamma^2}{(\gamma/2+\kappa)^4}\,\intensity \quad\text{(isolated mode)}.
\end{equation}
This expression reduces to Eq.~(\ref{eq:ecrsingle}) in the limit of vanishing radiation loss. 

Finally, we investigate the two-ring system again, but not at the EP. Instead, we optimize the intercavity coupling strength~$|V|$ such that, for given $\kappa$ and $\gamma$, the reduced QFI in Eq.~(\ref{eq:ecr1glossesred}) is largest. A short calculation shows that this happens for 
\begin{equation}\label{eq:Vopt}
|V|^2_{\text{opt}} = \frac{(\gamma+\kappa)\kappa}{4} . 
\end{equation}
Inserting this value into Eq.~(\ref{eq:ecr1glossesred}) yields for the two-ring setup with optimized intercavity coupling strength
\begin{equation}\label{eq:optlosses}
	\FIred_\varepsilon = \frac{4}{\kappa^2}\frac{\gamma^2}{(\gamma+\kappa)^2}\,\intensity \quad\text{(opt)}.
\end{equation}
In the limit of vanishing internal losses, the optimal intercavity coupling~(\ref{eq:Vopt}) also vanishes, while the reduced QFI~(\ref{eq:optlosses}) diverges. This behavior aligns with the lossless case presented in Eq.~(\ref{eq:ecr0}). 

\begin{figure}[ht]
\centering\includegraphics[width=0.75\columnwidth]{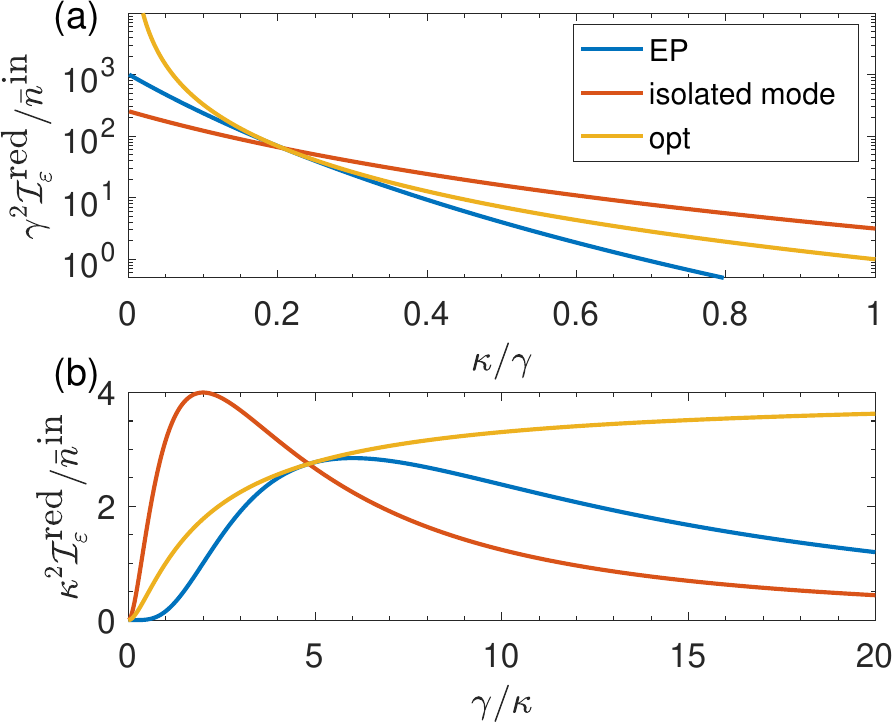}
\caption{The reduced QFI as function of (a) the internal losses~$\kappa$ normalized by the waveguide coupling strength~$\gamma$ and (b) $\gamma$ normalized by $\kappa$. Note the different scaling applied to the respective QFI to render it dimensionless. The two-ring setup at the EP [Eq.~(\ref{eq:ecr1lossesredEP})] is shown in blue, the single-ring setup with the isolated mode [Eq.~(\ref{eq:ecrsinglelossesred})] is shown in red, and the two-ring setup with optimized intercavity coupling strength [Eq.~(\ref{eq:optlosses})] is shown in yellow.}
\label{fig:losses}
\end{figure}

The reduced QFI for these three systems is presented in Fig.~\ref{fig:losses}, scaled in two different ways to render it dimensionless. In subfigure~(a), it is scaled by $\gamma^2/\intensity$, which is the natural choice when considering the dependence on the internal losses $\kappa$ (for fixed $\gamma$) coming from the case $\kappa = 0$ as described in Eqs.~(\ref{eq:ecr1}) and~(\ref{eq:ecrsingle}). The reduced QFI shows in all cases a monotonic decay as function of $\kappa/\gamma$. It is evident that the EP maintains its advantage over the isolated mode up to the critical value $\kappa_{\text{c}} = (\sqrt{2}-1)\gamma/2\approx 0.2\gamma$. Beyond this threshold, the isolated mode shows a larger reduced QFI. One reason is that as internal losses increase, the higher-order terms in the expansion of the Green's function in Eq.~(\ref{eq:GKato}) become progressively less dominant, as the pole moves deeper into the complex plane, moving farther away from the real axis. Another reason is that, in the single-ring setup, the information source is directly positioned in the ring closest to the waveguide, making it more easily accessible, in particular in the regime of large internal losses. 
The result of the optimized system is represented by the yellow curve. It is consistently above the result for the EP and touches the EP curve at $\kappa_{\text{c}}$. For $\kappa > \kappa_{\text{c}}$ it is below the result for the isolated mode.

Figure~\ref{fig:losses}(b) uses a scaling with $\kappa^2/\intensity$, which is the natural choice when considering the dependence on $\gamma$ (for fixed $\kappa$). Using this scaling, the reduced QFI exhibits a non-monotonic behavior as function of $\kappa/\gamma$ with a local maximum both for the isolated mode and for the EP. Note that the existence of this maximum is not an artefact of the scaling but appears naturally when we plot Eqs.~(\ref{eq:ecr1lossesredEP}), (\ref{eq:ecrsinglelossesred}), and~(\ref{eq:optlosses}) as function of $\gamma/\kappa$ instead of $\kappa/\gamma$.
In the isolated-mode case, the maximum occurs at critical coupling~\cite{CPV00}, $\gamma = 2\kappa$, which might be expected since this condition ensures the most efficient coupling between the waveguide and the information source located in the only ring. 
In the EP case, the cavity system is more complex; consequently, critical coupling to the upper ring alone does not determine the local maximum of the reduced QFI, which instead occurs at a significantly higher value of the waveguide coupling $\gamma = 6\kappa$. 

The contrasting regions in Fig.~\ref{fig:losses} characterized by small versus large internal losses partly explain the diverse findings in the literature concerning EP-enhancement, with some studies reporting its presence~\cite{ZSH19,ASF23,CKW26}, while others do not~\cite{LC18,CJL19,DMA22,AAK25}.
It is mentioned that using as information source the coupling of the two rings as implemented in Refs.~\cite{LS24,DMA22,AAK25} yields, in the case of no loss (see Sec.~\ref{sec:microringEP2}) and uniform internal loss considered here, a vanishing reduced QFI.

\begin{figure}[ht]
\centering\includegraphics[width=0.75\columnwidth]{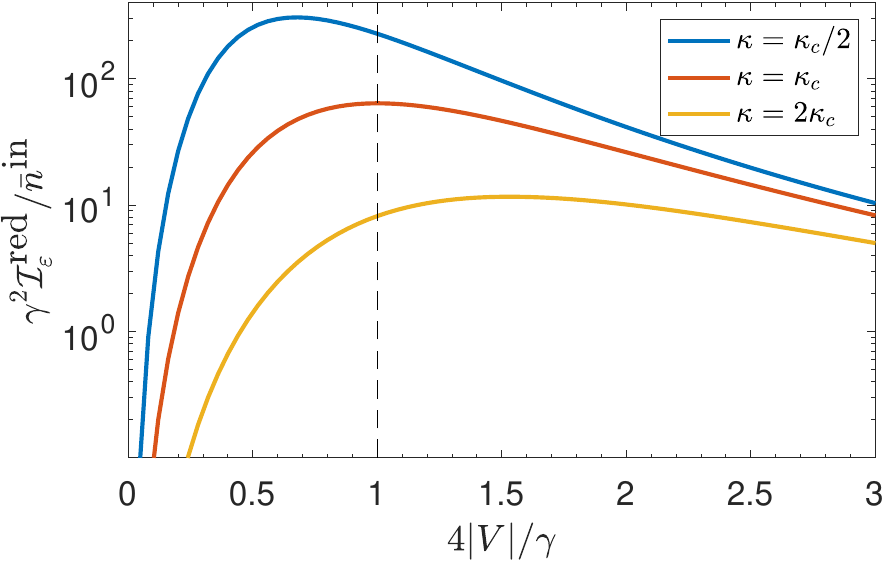}
\caption{The reduced QFI in Eq.~(\ref{eq:ecr1glossesred}) for the two-ring setup as function of the normalized intercavity coupling strength $|V|$ for three values of the internal losses~$\kappa$ (increasing from top to bottom); cf. the case of vanishing $\kappa$ in Fig.~\ref{fig:IvsV}. The EP is indicated by the dashed line.}
\label{fig:losses2}
\end{figure}

To investigate how the reduced QFI depends on the intercavity coupling strength $|V|$ in the presence of internal losses $\kappa$, we refer to Fig.~\ref{fig:losses2}. Unlike the case of vanishing internal losses shown in Fig.~\ref{fig:IvsV}, the QFI no longer diverges as $|V|$ vanishes. Instead, it exhibits a local maximum, whose position depends on $\kappa$. This maximum coincides with the EP only in the special case of critical internal losses. We emphasize once again that a missing peak of the reduced QFI at the EP does not imply that sensing at the EP is not enhanced when compared to the isolated mode. Still, of course, Figs.~\ref{fig:losses} and \ref{fig:losses2} also underline that being at the EP is not necessarily optimal for sensing applications.

\begin{figure}[ht]
\centering\includegraphics[width=0.75\columnwidth]{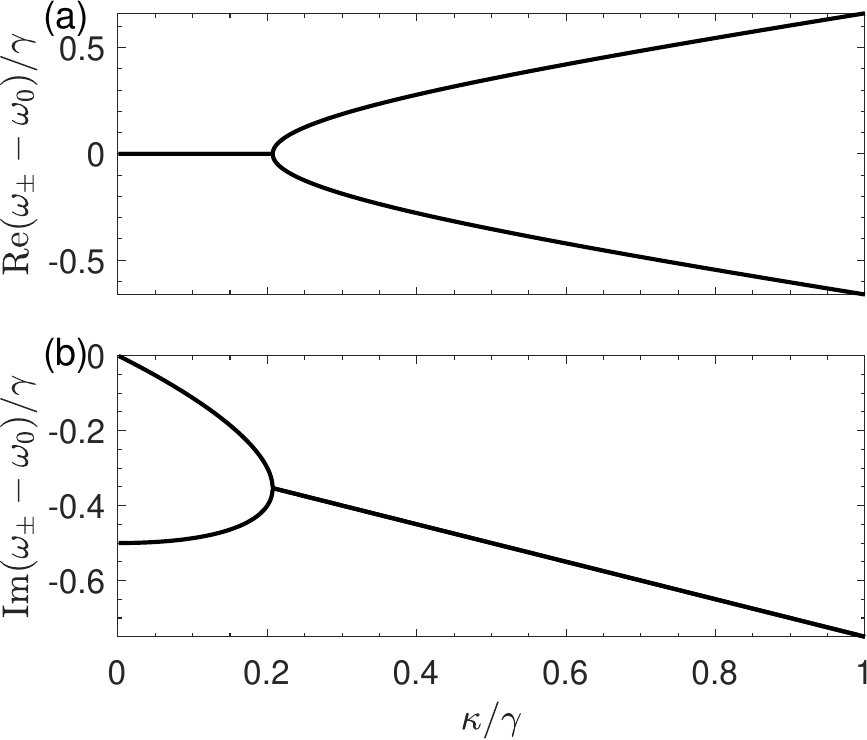}
\caption{The complex eigenfrequencies for the two-ring setup with optimized intercavity coupling strength [Eq.~(\ref{eq:Vopt})] as function of the internal losses $\kappa$; both axes are normalized to the waveguide coupling strength~$\gamma$. The real part is shown in (a) and the imaginary part in (b); cf. Fig.~\ref{fig:losses}(a).}
\label{fig:frequencies}
\end{figure}

To understand how the optimized system can outperform the EP case, we refer back to the lossless scenario examined in Sect.~\ref{sec:microringEP2}, where the linewidth splitting in the weak coupling regime is the critical mechanism. In the presence of small internal losses, a similar situation occurs, as illustrated in Fig.~\ref{fig:frequencies}. 
Again, there is a weak coupling regime with a splitting of the imaginary parts of the eigenfrequencies. The resonant phase shift associated with the smaller imaginary part is very narrow and thus very sensitive to system perturbations, thereby giving rise to more Fisher information at the output.

The weak coupling region persists up to the critical internal loss~$\kappa_{\text{c}}$ where an EP is reached, see Fig.~\ref{fig:frequencies}. Only at this value of $\kappa$ the EP is the optimal choice.
In the strong coupling regime $\kappa > \kappa_{\text{c}}$, we observe a splitting in the real part of the eigenfrequencies. In this regime, where internal losses are comparatively high, the intercavity coupling strength at the EP, $|V| = \gamma/4$, proves insufficient for coupling the incoming light to the information source. Consequently, the system compensates by increasing the intercavity coupling strength.

\section{Conclusion}
\label{sec:conclusion}
We have studied non-Hermitian sensor systems using a recently developed scattering-matrix approach to calculate the QFI for the scattering of coherent light. For the case without internal losses, we revealed that the maximal QFI per incoming photon flux is determined by three key factors: the decay rate of the resonant mode, the strength of the spectral response, and the adjustment between the scattering states and the information source. In particular, we have shown that the QFI is enhanced at an EP when compared to an isolated mode or a DP with equal decay rate. 
This has been made explicit by deriving upper bounds for the QFI in terms of the spectral response strength and decay rate.
For a localized information source, we found that the QFI per incoming photon flux depends exclusively on the LDOS, consistent with Ref.~\cite{BCM20}.

While the system at the EP outperforms the corresponding system at the isolated mode or DP, the QFI can be further increased by deviating from the EP due to the non-Hermitian phenomenon of linewidth splitting which reduces the decay rate of one mode. The QFI is diverging to infinity when a mode with vanishing decay rate is formed.  
This scenario remains valid for sufficiently small internal losses.  

Other studies report the absence of an enhancement of the QFI at the EP~\cite{LC18,CJL19,DMA22,LS24,AAK25}. This discrepancy can be attributed to the omission of at least one of the crucial aspects revealed in this work: the adjustment between the scattering states and the information source, the decay rate of the resonant mode, and the strength of the spectral response. In detail:
(i) The information source and the cavity system have not always been perfectly adjusted to each other. For instance, the selection of the information source in Refs.~\cite{DMA22,LS24,AAK25} is ineffective in the case of spatially uniform losses and, as a result, may not be optimal when losses are slightly nonuniform. 
(ii) The performance of the EP has not been compared to a corresponding isolated mode or DP, but rather to other systems exhibiting one mode with lower decay rate such as in Ref.~\cite{AAK25}.   
(iii) Additional noise channels to implement the EPs are needed in almost all of the above studies. This is avoided in our approach as we do not need to include additional noise channels except those inherent to the scattering setup. Moreover, the cavity systems consisting of two or more microrings attached to a waveguide are chosen such that the spectral response strength is maximal for given decay rate.


The argument from Refs.~\cite{DWC23,ZLX25}, suggesting that any non-Hermitian system can be embedded within a Hermitian one, is not applicable in our scattering setup, as the cavity system without the coupling to the scattering channels is inherently Hermitian. The non-Hermiticity present in the scattering system arises solely from the coupling to incoming and outgoing channels, and it is an unavoidable aspect of the scattering setup.

From a practical standpoint, the scattering-based analysis presented here translates into clear guidelines for maximizing the Fisher information in non-Hermitian sensing experiments. The achievable Fisher information per incoming photon flux is controlled by three interdependent factors: the decay rate of the resonant mode interacting with the perturbation, the degree of non-normality quantified by the spectral response strength, and the spatial and modal overlap between the information source and the scattering states. For localized perturbations, this dependence simplifies to a particularly transparent criterion: the Fisher information is directly determined by the local density of states at the perturbation site, favoring configurations in which the information source is placed at a field maximum of a long-lived mode. Exceptional points enhance sensitivity insofar as they increase non-normality at fixed decay rate; however, the optimal operating point is not necessarily located exactly at the EP. In practice, a slight detuning from the EP can further increase the Fisher information by inducing non-Hermitian linewidth splitting and thereby forming a mode with reduced decay rate and enhanced local density of states. This enhancement persists in the presence of sufficiently small internal losses, while larger losses ultimately favor simpler isolated resonances. 
\HL{An interesting direction for future work concerns the extension of the present framework to multi-parameter estimation problems~(see e.g.~\cite{NCK23}), where non-Hermitian systems may exhibit nontrivial tradeoffs between the simultaneous estimation of frequencies, decay rates, couplings, or spatially distributed perturbations. In such settings, the interplay between non-normality and the compatibility of optimal measurements may lead to additional constraints and opportunities beyond the single-parameter scenario considered here.}
Finally, reaching the fundamental sensitivity bound requires appropriate matching of the input wavefront to the sensing channel and phase-sensitive detection, as the Fisher information is often encoded predominantly in the phase response rather than in intensity alone.


\ack{We thank Jakob H{\"u}pfl for helpful discussions.}

\funding{Funded by the Deutsche Forschungsgemeinschaft (DFG, German Research Foundation) [WI 1986/14-1] and by the Austrian Science Fund (FWF) [10.55776/PIN7240924]. The authors acknowledge TU Wien Bibliothek for financial support through its Open Access Funding Program.} 

\HL{
\section*{Data availability statement}
Supplementary source code for the numerical experiments for this work are openly available at the following DOI: https://doi.org/10.5281/zenodo.18864039

\section*{Author contributions}

J.W. conceived the idea. J.W. and S.R. developed the theory and carried out the analytical and numerical investigations. J.W. and S.R. wrote and revised the manuscript. All authors contributed to analysis and discussion of the results.

\section*{Conflict of interest}

The authors declare no competing interests.
}

\bibliographystyle{unsrt}

\begin{thebibliography}{10}

\bibitem{Kato66}
T.~Kato.
\newblock {\em Perturbation Theory for Linear Operators}.
\newblock Springer, New York, 1966.

\bibitem{Heiss00}
W.~D. Heiss.
\newblock Repulsion of resonance states and exceptional points.
\newblock {\em Phys. Rev. E}, 61:929--932, 2000.

\bibitem{Berry04}
M.~V. Berry.
\newblock Physics of nonhermitian degeneracies.
\newblock {\em Czech. J. Phys.}, 54:1039--1047, 2004.

\bibitem{Heiss04}
W.~D. Heiss.
\newblock Exceptional points of non-{H}ermitian operators.
\newblock {\em J. Phys. A: Math. Gen.}, 37:2455--2464, 2004.

\bibitem{MA19}
M.-A. Miri and A.~Al\`{u}.
\newblock Exceptional points in optics and photonics.
\newblock {\em Science}, 363(6422):eaar7709, 2019.

\bibitem{POR14}
B.~Peng, {\c{S}}.~K. {\"O}zdemir, S.~Rotter, H.~Y{\i}lmaz, M.~Liertzer,
  F.~Monifi, C.~M. Bender, F.~Nori, and L.~Yang.
\newblock Loss-induced suppression and revival of lasing.
\newblock {\em Science}, 17:328--332, 2014.

\bibitem{HMH14}
H.~Hodaei, M.-A. Miri, M.~Heinrich, D.~Christodoulides, and M.~Khajavikhan.
\newblock Parity-time-symmetric microring lasers.
\newblock {\em Science}, 346:975--979, 2014.

\bibitem{MZS16}
P.~Miao, Z.~Zhang, J.~Sun, W.~Walasik, S.~Longhi, N.~M. Litchinitser, and
  L.~Feng.
\newblock Orbital angular momentum microlaser.
\newblock {\em Science}, 353:464--467, 2016.

\bibitem{POL16}
B.~Peng, {\c{S}}.~K. {\"O}zdemir, M.~Liertzer, W.~Chen, J.~Kramer, H.~Yilmaz,
  J.~Wiersig, S.~Rotter, and L.~Yang.
\newblock Chiral modes and directional lasing at exceptional points.
\newblock {\em Proc. Natl. Acad. Sci. USA}, 113:6845--6850, 2016.

\bibitem{XMJ16}
H.~Xu, D.~Mason, L.~Jiang, and J.~G.~E Harris.
\newblock Topological energy transfer in an optomechanical system with
  exceptional points.
\newblock {\em Nature (London)}, 537:80--85, 2016.

\bibitem{DMB16}
J.~Doppler, A.~A. Mailybaev, J.~B{\"o}hm, U.~Kuhl, A.~Girschik, F.~Libisch,
  T.~J. Milburn, P.~Rabl, N.~Moiseyev, and S.~Rotter.
\newblock Dynamically encircling an exceptional point for asymmetric mode
  switching.
\newblock {\em Nature (London)}, 537:76--79, 2016.

\bibitem{RMS17}
S.~Richter, T.~Michalsky, C.~Sturm, B.~Rosenow, M.~Grundmann, and
  R.~Schmidt-Grund.
\newblock Exceptional points in anisotropic planar microcavities.
\newblock {\em Phys. Rev. A}, 95:023836, 2017.

\bibitem{SHR19}
W.~R. Sweeney, C.~W. Hsu, S.~Rotter, and A.~D. Stone.
\newblock Perfectly absorbing exceptional points and chiral absorbers.
\newblock {\em Phys. Rev. Lett}, 122:093901, 2019.

\bibitem{ZOE20}
Q.~Zhong, {\c{S}}.~K. {\"O}zdemir, A.~Eisfeld, A.~Metelmann, and R.~El-Ganainy.
\newblock Exceptional points-based optical amplifiers.
\newblock {\em Phys. Rev. Appl.}, 13:014070, 2020.

\bibitem{XLJ23}
Y.~Xu, L.~Li, H.~Jeong, S.~Kim, I.~Kim, J.~Rho, and Y.~Liu.
\newblock Subwavelength control of light transport at the exceptional point by
  non-{H}ermitian metagratings.
\newblock {\em Sci. Adv.}, 9:eadf3510, 2023.

\bibitem{YZZ23}
M.~Yang, L.~Zhu, Q.~Zhong, R.~El-Ganainy, and P.-Y. Chen.
\newblock Spectral sensitivity near exceptional points as a resource for
  hardware encryption.
\newblock {\em Nat. Commun.}, 14:1145, 2023.

\bibitem{Wiersig14b}
J.~Wiersig.
\newblock Enhancing the {S}ensitivity of {F}requency and {E}nergy {S}plitting
  {D}etection by {U}sing {E}xceptional {P}oints: {A}pplication to {M}icrocavity
  {S}ensors for {S}ingle-{P}article {D}etection.
\newblock {\em Phys. Rev. Lett.}, 112:203901, 2014.

\bibitem{Wiersig16}
J.~Wiersig.
\newblock Sensors operating at exceptional points: General theory.
\newblock {\em Phys. Rev. A}, 93:033809, 2016.

\bibitem{Wiersig22}
J.~Wiersig.
\newblock Response strengths of open systems at exceptional points.
\newblock {\em Phys. Rev. Res.}, 4:023121, 2022.

\bibitem{Wiersig23b}
J.~Wiersig.
\newblock Moving along an exceptional surface towards a higher-order
  exceptional point.
\newblock {\em Phys. Rev. A}, 108:033501, 2023.

\bibitem{COZ17}
W.~Chen, {\c{S}}.~K. {\"O}zdemir, G.~Zhao, J.~Wiersig, and L.~Yang.
\newblock Exceptional points enhance sensing in an optical microcavity.
\newblock {\em Nature (London)}, 548:192--196, 2017.

\bibitem{HHW17}
H.~Hodaei, A.~Hassan, S.~Wittek, H.~Carcia-Cracia, R.~El-Ganainy,
  D.~Christodoulides, and M.~Khajavikhan.
\newblock Enhanced sensitivity at higher-order exceptional points.
\newblock {\em Nature (London)}, 548:187--191, 2017.

\bibitem{XLK19}
Z.~Xiao, H.~Li, T.~Kottos, and A.~Al{\`u}.
\newblock Enhanced sensing and nondegraded thermal noise performance based on
  {PT}-symmetric electronic circuits with a sixth-order exceptional point.
\newblock {\em Phys. Rev. Lett.}, 123:213901, 2019.

\bibitem{LLS19}
Y.-H. Lai, Y.-K. Lu, M.-G. Suh, Z.~Yuan, and K.~Vahala.
\newblock Observation of the exceptional-point-enhanced {S}agnac effect.
\newblock {\em Nature (London)}, 576:65--69, 2019.

\bibitem{HSC19}
M.~P. Hokmabadi, A.~Schumer, D.~N. Christodoulides, and M.~Khajavikhan.
\newblock Non-{H}ermitian ring laser gyroscopes with enhanced {S}agnac
  sensitivity.
\newblock {\em Nature (London)}, 576:70--74, 2019.

\bibitem{PNC20}
J.-H. Park, A.~Ndao, L.~Hsu, A.~Kodigala, T.~Lepetit, Y.-H. Lo, and B.~Kant\'e.
\newblock Symmetry-breaking-induced plasmonic exceptional points and nanoscale
  sensing.
\newblock {\em Nat. Physics}, 16:462--468, 2020.

\bibitem{KCE22}
R.~Kononchuk, J.~Cai, F.~Ellis, R.~Thevamaran, and T.~Kottos.
\newblock Exceptional-point-based accelerometers with enhanced signal-to-noise
  ratio.
\newblock {\em Nature (London)}, 607:697--702, 2022.

\bibitem{DLY19}
Z.~Dong, Z.~Li, F.~Yang, C.-W. Qiu, and J.~S. Ho.
\newblock Sensitive readout of implantable microsensors using a wireless system
  locked to an exceptional point.
\newblock {\em Nat. Electron.}, 2:335--342, 2019.

\bibitem{LCL23}
Z.. Li, J.~Chen, L.~Li, J.~Zhang, and J.~Yao.
\newblock Exceptional-point-enhanced sensing in an all-fiber bending sensor.
\newblock {\em Opto-Electron Adv}, 6:230019, 2023.

\bibitem{DDS25}
Y.~Dong, L.~J. Dong, and X.~Y. Shen.
\newblock A study on the detection of weathering in the {Y}ungang {G}rottoes
  using non-{H}ermitian wireless sensing.
\newblock {\em AIP Advances}, 15:075021, 2025.

\bibitem{YZY25}
Z.~Ye, G.~Zhao, M.~Yang, Y.~Xu, Y.~Ren, Z.~Chen, S.~M. Andrabi, J.~Xie, W.~Gad,
  Z.~Yan, and P.-Y. Chen.
\newblock A highly sensitive and multiplexed wireless sensing system with
  skin-like compliance and stretchability for wearable applications.
\newblock {\em Sci. Adv.}, 11:eadt4923, 2025.

\bibitem{Wiersig20b}
J.~Wiersig.
\newblock Prospects and fundamental limits in exceptional point-based sensing.
\newblock {\em Nat. Commun.}, 11:2454, 2020.

\bibitem{Wiersig20c}
J.~Wiersig.
\newblock Review of exceptional point-based sensors.
\newblock {\em Photonics Res.}, 8:1457--1467, 2020.

\bibitem{WLY20}
H.~Wang, Y.-H. Lai, Z.~Yuan, M.-G. Suh, and K.~Vahala.
\newblock Petermann-factor sensitivity limit near an exceptional point in a
  {B}rillouin ring laser gyroscope.
\newblock {\em Nat. Commun.}, 11:1610, 2020.

\bibitem{XML24}
J.~Xu, Y.~Mao, Z.~Li, Y.~Zuo, J.~Zhang, B.~Yang, W.~Xu, N.~Liu, Z.~J. Deng,
  W.~Chen, K.~Xia, C.-W. Qiu, Z.~Zhu, H.~Jing, and K.~Liu.
\newblock Single-cavity loss-enabled nanometrology.
\newblock {\em Nature Nanotechnology}, 19:1472--1477, 2024.

\bibitem{Langbein18}
W.~Langbein.
\newblock No exceptional precision of exceptional-point sensors.
\newblock {\em Phys. Rev. A}, 98:023805, 2018.

\bibitem{LC18}
H.-K. Lau and A.~A. Clerk.
\newblock Fundamental limits and non-reciprocal approaches in non-{H}ermitian
  quantum sensing.
\newblock {\em Nat. Commun.}, 9:4320, 2018.

\bibitem{ZSH19}
M.~Zhang, W.~Sweeney, C.~W. Hsu, L.~Yang, A.~D. Stone, and L.~Jiang.
\newblock Quantum noise theory of exceptional point amplifying sensors.
\newblock {\em Phys. Rev. Lett.}, 123:180501, 2019.

\bibitem{CJL19}
C.~Chen, L.~Jin, and R.-B. Liu.
\newblock Sensitivity of parameter estimation near the exceptional point of a
  non-{H}ermitian system.
\newblock {\em New J. Phys.}, 21:083002, 2019.

\bibitem{DMA22}
R.~Duggan, S.~A. Mann, and A~Al{\`u}.
\newblock Limitations of sensing at an exceptional point.
\newblock {\em ACS Photonics}, 9:1554--1566, 2022.

\bibitem{Sunada18}
S.~Sunada.
\newblock Enhanced response of non-{H}ermitian photonic systems near
  exceptional points.
\newblock {\em Phys. Rev. A}, 97:043804, 2018.

\bibitem{BC94}
S.~L. Braunstein and C.~M. Caves.
\newblock Statistical distance and the geometry of quantum states.
\newblock {\em Phys. Rev. Lett.}, 72:3439, 1994.

\bibitem{BG13}
D.~C. Brody and E.-M. Graefe.
\newblock Information geometry of complex {H}amiltonians and exceptional
  points.
\newblock {\em Entropy}, 15:3361--3378, 2013.

\bibitem{AAK25}
A.~Almanakly, R.~Assouly, H.H. Kang, M.~Gingras, B.~M. Niedzielski,
  H.~Stickler, M.~E. Schwartz, K.~Serniak, M.~Hays, J.~A. Grover, and W.~D.
  Oliver.
\newblock Probing sensitivity near a quantum exceptional point using waveguide
  quantum electrodynamics.
\newblock {\em arXiv:2510.21554v1}, 2025.

\bibitem{ASF23}
D.~Anderson, M.~Shah, and L.~Fan.
\newblock Clarification of the exceptional-point contribution to photonic
  sensing.
\newblock {\em Phys. Rev. Appl.}, 19:034059, 2023.

\bibitem{NCK23}
J.~Naikoo, R.~W. Chhajlany, and J.~Ko\l{}ody\'{n}ski.
\newblock Multiparameter estimation perspective on non-{H}ermitian
  singularity-enhanced sensing.
\newblock {\em Phys. Rev. Lett.}, 131:220801, 2023.

\bibitem{CKW26}
R.~L. Cook, L.~Ko, and B.~Whaley.
\newblock Optimal sensing on an asymmetric exceptional surface.
\newblock {\em Phys. Rev. Res.}, 8:013031, 2026.

\bibitem{DWC23}
W.~Ding, X.~Wang, and S.~Chen.
\newblock Fundamental {S}ensitivity {L}imits for {N}on-{H}ermitian {Q}uantum
  {S}ensors.
\newblock {\em Phys. Rev. Lett}, 131:160801, 2023.

\bibitem{ZLX25}
N.~Zeng, T.~Liu, K.~Xia, Y.-R. Zhang, and F.~Nori.
\newblock Non-{H}ermitian sensing from the perspective of post-selected
  measurements.
\newblock {\em Phys. Rev. Research}, 7:043219, 2025.

\bibitem{ANO26}
I.~I. Arkhipov, F.~Nori, and {\c{S}}.~K. {\"O}zdemir.
\newblock Achieving the quantum {F}isher information bound in
  pseudo-{H}ermitian sensors.
\newblock {\em Phys. Rev. Lett.}, 136:080802, 2026.

\bibitem{ZC25}
Y.~D. Zheng, X.and~Chong.
\newblock Noise constraints for nonlinear exceptional point sensing.
\newblock {\em Phys. Rev. Lett.}, 134:133801, 2025.

\bibitem{BRM21}
D.~Bouchet, S.~Rotter, and A.~P. Mosk.
\newblock Maximum information states for coherent scattering measurements.
\newblock {\em Nat. Phys.}, 17:564--568, 2021.

\bibitem{PLC10}
S.~M. Popoff, G.~Lerosey, R.~Carminati, M.~Fink, A.~C. Boccara, and S.~Gigan.
\newblock Measuring the transmission matrix in optics: An approach to the study
  and control of light propagation in disordered media.
\newblock {\em Phys. Rev. Lett.}, 104:100601, 2010.

\bibitem{RG17}
S.~Rotter and S.~Gigan.
\newblock Light fields in complex media: {M}esoscopic scattering meets wave
  control.
\newblock {\em Rev. Mod. Phys.}, 89:015005, 2017.

\bibitem{CMR22}
H.~Cao, A.~P. Mosk, and S.~Rotter.
\newblock Shaping the propagation of light in complex media.
\newblock {\em Nat. Phys.}, 18:994--1007, 2022.

\bibitem{HRR24}
J.~H{\"u}pfl, F.~Russo, L.~M. Rachbauer, D.~Bouchet, J.~Lu, U.~Kuhl, and
  S.~Rotter.
\newblock Continuity equation for the flow of {F}isher information in wave
  scattering.
\newblock {\em Nat. Phys.}, 20:1294--1299, 2024.

\bibitem{Johnston21}
N.~Johnston.
\newblock {\em Advanced Linear and Matrix Algebra}.
\newblock Springer, Switzerland, 2021.

\bibitem{Wiersig22b}
J.~Wiersig.
\newblock Distance between exceptional points and diabolic points and its
  implication for the response strength of non-{H}ermitian systems.
\newblock {\em Phys. Rev. Res.}, 4:033179, 2022.

\bibitem{Wiersig22c}
J.~Wiersig.
\newblock Revisiting the hierarchical construction of higher-order exceptional
  points.
\newblock {\em Phys. Rev. A}, 106:063526, 2022.

\bibitem{Wiersig23}
J.~Wiersig.
\newblock Petermann factors and phase rigidities near exceptional points.
\newblock {\em Phys. Rev. Res.}, 5:033042, 2023.

\bibitem{KWS25}
J.~Kullig, J.~Wiersig, and H.~Schomerus.
\newblock Generalized {P}etermann factor of non-{H}ermitian systems at
  exceptional points.
\newblock {\em Phys. Rev. Res.}, 7:043246, 2025.

\bibitem{KW25}
J.~Kullig and J.~Wiersig.
\newblock Calculating the spectral response strength of non-{H}ermitian systems
  with an exceptional point directly from wave simulations.
\newblock {\em Phys. Rev. Res.}, 7:013223, 2025.

\bibitem{HKB20}
M.~Horodynski, M.and~K{\"uh}mayer, A.~Brandst{\"o}tter, K.~Pichler, Y.~V.
  Fyodorov, U.~Kuhl, and S.~Rotter.
\newblock Optimal wave fields for micromanipulation in complex scattering
  environments.
\newblock {\em Nat. Photonics}, 14:149--153, 2020.

\bibitem{MW69}
C.~Mahaux and H.~A. Weidenm{\"u}ller.
\newblock {\em Shell-model approach to nuclear reactions}.
\newblock North-Holland, Amsterdam, 1969.

\bibitem{ABB17}
P.~Ambichl, A.~Brandst{\"o}tter, J.~B{\"o}hm, M.~K{\"uh}mayer, U.~Kuhl, and
  S.~Rotter.
\newblock Focusing inside disordered media with the generalized
  {W}igner-{S}mith operator.
\newblock {\em Phys. Rev. Lett.}, 119:033903, 2017.

\bibitem{GCM96}
V.~Gasparian, T.~Christen, and M.~B{\"u}ttiker.
\newblock Partial densities of states, scattering matrices, and {G}reen’s
  functions.
\newblock {\em Phys. Rev. A}, 54(5):4022--4031, 1996.

\bibitem{HBK21}
M.~Horodynski, D.~Bouchet, M.~K{\"uh}mayer, and S.~Rotter.
\newblock Invariance {P}roperty of the {F}isher {I}nformation in {S}cattering
  {M}edia.
\newblock {\em Phys. Rev. Lett.}, 127:233201, 2021.

\bibitem{BCM20}
D.~Bouchet, R.~Carminati, and A.~P. Mosk.
\newblock Influence of the local scattering environment on the localization
  precision of single particles.
\newblock {\em Phys. Rev. Lett.}, 124:133903, 2020.

\bibitem{LPL16}
Z.~Lin, A.~Pick, M.~Loncar, and A.~W. Rodriguez.
\newblock Enhanced spontaneous emission at third-order {D}irac exceptional
  points in inverse-designed photonic crystals.
\newblock {\em Phys. Rev. Lett.}, 117:107402, 2016.

\bibitem{PZM17}
A.~Pick, B.~Zhen, O.~D. Miller, C.~W. Hsu, F.~Herandez, A.~W. Rodriguez,
  M.~Solja\u{c}i\'c, and S.~G. Johnson.
\newblock General theory of spontaneous emission near exceptional points.
\newblock {\em Opt. Express}, 25:12325, 2017.

\bibitem{HJ13}
R.~A. Horn and C.~R. Johnson.
\newblock {\em Matrix Analysis}.
\newblock Cambridge University Press, Cambridge, 2013.

\bibitem{LS24}
H.~Loughlin and V.~Sudhir.
\newblock Exceptional-point sensors offer no fundamental signal-to-noise ratio
  enhancement.
\newblock {\em Phys. Rev. Lett.}, 132:243601, 2024.

\bibitem{FAB25}
B.~Franchi, R.and~Aslan, S.~Biasi, and L.~Pavesi.
\newblock Controlled comparison of phase-tuned exceptional and diabolic point
  sensing in integrated optical microresonators.
\newblock {\em Appl. Phys. Lett.}, 127:201103, 2025.

\bibitem{FriWin85}
Harald Friedrich and Dieter Wintgen.
\newblock Interfering resonances and bound states in the continuum.
\newblock {\em Phys. Rev. A}, 32(6):3231--3242, 1985.

\bibitem{Wiersig06}
J.~Wiersig.
\newblock Formation of long-lived, scarlike modes near avoided resonance
  crossings in optical microcavities.
\newblock {\em Phys. Rev. Lett.}, 97:253901, 2006.

\bibitem{RotterNC14}
M.~Brandstetter, M.~Liertzer, C.~Deutsch, P.~Klang, J.~Sch{\"o}berl, H.~E.
  T{\"u}reci, G.~Strasser, K.~Unterrainer, and S.~Rotter.
\newblock Reversing the pump dependence of a laser at an exceptional point.
\newblock {\em Nat. Commun.}, 5:4034, 2014.

\bibitem{KZW25}
J.~Kullig, Q.~Zhong, J.~Wiersig, and R.~El-Ganainy.
\newblock Exceptional {P}oints and {L}asing {T}hresholds: {W}hen {L}ower-{Q}
  {M}odes {W}in.
\newblock {\em Phys. Rev. Lett.}, 135:173802, 2025.

\bibitem{SAC06}
A.~V. Sokolov, A.~A. Andrianov, and F.~Cannata.
\newblock Non-{H}ermitian quantum mechanics of non-diagonalizable
  {H}amiltonians: puzzles with self-orthogonal states.
\newblock {\em J. Phys. A: Math. Gen.}, 39:10207--10227, 2006.

\bibitem{YSS11}
G.~Yoo, H.-S. Sim, and H.~Schomerus.
\newblock Quantum noise and mode nonorthogonality in non-{H}ermitian
  {PT}-symmetric optical resonators.
\newblock {\em Phys. Rev. A}, 84:063833, 2011.

\bibitem{WLX24}
C.~Wang, N.~Li, J.~Xie, C.~Ding, Z.~Ji, L.~Xiao, S.~Jia, Y.~Hu, and Y.~Zhao.
\newblock Exceptional {N}exus in {B}ose-{E}instein {C}ondensates with
  {C}ollective {D}issipation.
\newblock {\em Phys. Rev. Lett.}, 132:253401, 2024.

\bibitem{ZRK19}
Q.~Zhong, J.~Ren, M.~Khajavikhan, D.~N. Christodoulides, {\c{S}}.~K.
  {\"O}zdemir, and R.~El-Ganainy.
\newblock Sensing with {E}xceptional {S}urfaces in {O}rder to {C}ombine
  {S}ensitivity with {R}obustness.
\newblock {\em Phys. Rev. Lett.}, 122:153902, 2019.

\bibitem{SZM22}
S.~Soleymani, Q.~Zhong, M.~Mokim, S.~Rotter, R.~El-Ganainy, and {\c{S}}.~K.
  {\"O}zdemir.
\newblock Chiral and degenerate perfect absorption on exceptional surfaces.
\newblock {\em Nat. Commun.}, 13:599, 2022.

\bibitem{Wiersig18b}
J.~Wiersig.
\newblock Non-{H}ermitian effects due to asymmetric backscattering of light in
  whispering-gallery microcavities.
\newblock In D.~Christodoulides and J.~Yang, editors, {\em Parity-time Symmetry
  and Its Applications}, pages 155--184. Springer, Singapore, 2018.

\bibitem{WES11}
J.~Wiersig, A.~Ebersp{\"a}cher, J.-B Shim, J.-W. Ryu, S.~Shinohara,
  M.~Hentschel, and H.~Schomerus.
\newblock Nonorthogonal pairs of copropagating optical modes in deformed
  microdisk cavities.
\newblock {\em Phys. Rev. A}, 84:023845, 2011.

\bibitem{Wiersig11}
J.~Wiersig.
\newblock Structure of whispering-gallery modes in optical microdisks perturbed
  by nanoparticles.
\newblock {\em Phys. Rev. A}, 84:063828, 2011.

\bibitem{CPV00}
M.~Cai, O.~Painter, and K.~J. Vahala.
\newblock Observation of critical coupling in a fiber taper to a
  silica-microsphere whispering-gallery mode system.
\newblock {\em Phys. Rev. Lett.}, 85:74, 2000.

\end{thebibliography}

\end{document}